# A Framework for Understanding the Impact of Integrating Conceptual and Quantitative Reasoning in a Quantum Optics Tutorial on Students' Conceptual Understanding

**Paul D. Justice** [1,*], **Emily Marshman and Chandralekha Singh** [2]

[1] Department of Physics, University of Cincinnati, Cincinnati, OH 45221, USA;
[2] Department of Physics, Community College of Allegheny County, Pittsburgh, PA 15212;
[3] Department of Physics, University of Pittsburgh, Pittsburgh, PA 15260;
* Correspondence: justicpl@ucmail.uc.edu
[2]emarshman@ccac.edu (E.M.); [3] clsingh@pitt.edu (C.S.)

**Abstract**

We investigated the impact of incorporating quantitative reasoning for deeper sense-making in a Quantum Interactive Learning Tutorial (QuILT) on students' conceptual performance using a framework emphasizing integration of conceptual and quantitative aspects of quantum optics. In this investigation, we compared two versions of the QuILT that were developed and validated to help students learn various aspects of quantum optics using a Mach Zehnder Interferometer with single photons and polarizers. One version of the QuILT is entirely conceptual while the other version integrates quantitative and conceptual reasoning (hybrid version). Performance on conceptual questions of upper-level undergraduate and graduate students who engaged with the hybrid QuILT was compared with that of those who utilized the conceptual QuILT emphasizing the same concepts. Both versions of the QuILT focus on the same concepts, use a scaffolded approach to learning, and take advantage of research on students' difficulties in learning these challenging concepts as well as a cognitive task analysis from an expert perspective as a guide. The hybrid and conceptual QuILTs were used in courses for upper-level undergraduates or first-year physics graduate students in several consecutive years at the same university. The same conceptual pre-test and post-test were administered after traditional lecture-based instruction in relevant concepts and after student engaged with the QuILT, respectively. We find that the post-test performance of physics graduate students who utilized the hybrid QuILT on conceptual questions, on average, was better than those who utilized the conceptual QuILT. For undergraduates, the results showed differences for different classes. One possible interpretation of these findings that is consistent with our framework is that integrating conceptual and quantitative aspects of physics in research-based tools and pedagogies should be commensurate with students' prior knowledge of physics and mathematics involved so that students do not experience cognitive overload while engaging with such learning tools and have appropriate opportunities for metacognition, deeper sense-making, and knowledge organization. In the undergraduate course in which many students did not derive added benefit from the integration of conceptual and quantitative aspects, their pre-test performance suggests that the traditional lecture-based instruction may not have sufficiently provided a "first coat" to help students avoid cognitive overload when engaging with the hybrid QuILT. These findings suggest that different groups of students can benefit from a research-based learning tool that integrates conceptual and quantitative aspects if cognitive overload while learning is prevented either due to students' high mathematical facility or due to their reasonable conceptual facility before engaging with the learning tool.

## 1. Introduction, Framework, and Goals

*1.1. Background on Expertise in Physics*

To help students develop expertise in any area of physics [1], one must first ask how experts, in general, compare to novices in terms of their knowledge organization and their problem-solving, reasoning, and metacognitive skills [2-7]. Metacognitive skills refer to a set of tools and activities that can help individuals control their learning [8, 9]. Schoenfeld [9] emphasized that sense-making requires metacognitive processes and that developing expertise and "learning to think mathematically means (a) developing a mathematical point of view—valuing the processes of mathematization and abstraction and having the predilection to apply them, and (b) developing competence with the tools of the trade and using those tools in the service of the goal of understanding structure—mathematical sense-making". The three main metacognitive skills are planning, self-monitoring, and evaluation [8, 9].

If our goal is to help students become experts in physics at any level, we must contemplate whether there is something special about the nature of expertise in physics over and above what we know about expertise in other disciplines, e.g., what is needed for becoming an expert tennis or chess player or music performer [10-12]. Physics expertise involves several key characteristics that distinguish experts from novices. Experts organize their knowledge hierarchically around fundamental principles rather than surface features, allowing them to recognize deep structural similarities between problems that may appear different superficially [13]. They possess both extensive content knowledge and sophisticated procedural knowledge (e.g., problem-solving strategies, mathematical techniques), with these elements integrated into coherent mental schemas. Experts often are good at identifying relevant principles and constraints in a problem situation, while novices often focus on superficial features and also have alternative conceptions. Additionally, experts demonstrate strong metacognitive awareness, continuously monitoring their understanding and adjusting their approach when needed.

Physics is a discipline that focuses on unraveling the underlying mechanisms of physical phenomena in our universe. Physicists engage in sense-making that involves an intricate coupling of physics and mathematics to build and refine models to test and explain physical phenomena that are observed or to predict those that have not been observed so far [14-35]. Uhden et al. [19] nicely elaborate upon the difference between technical and structural roles of mathematics in physics and that "the technical skills are associated with pure mathematical manipulations whereas the structural skills are related to the capacity of employing mathematical knowledge for structuring physical situations". Tzanakis' view [27] that "mathematics is the language of physics, not only as a tool for expressing, handling and developing logically physical concepts, methods and theories, but also as an indispensable, formative characteristic that shapes them, by deepening, sharpening, and extending their meaning, or even endowing them with meaning" is consistent with the view by Uhden et al. [19]. To help students develop physics expertise, it is important to recognize that there are very few fundamental physical laws which are encapsulated in compact mathematical forms and learning to unpack them can help one develop expertise and organize one's knowledge hierarchically (e.g., see [15, 19, 20, 23, 24, 27, 29-31, 34]). Expert-like sense-making of physical phenomena using cohesive physical models requires intricate coupling and synthesis of both conceptual and quantitative knowledge as shown schematically in Figure 1. Therefore, an important aspect of expertise in physics is the proficiency and fluidity with which one can seamlessly move back and forth and make appropriate connections between physical and mathematical concepts necessary to understand physical phenomena. In particular, developing expertise in physics entails making appropriate math–physics connections to meaningfully unpack, interpret, and apply the laws of physics and to use this sense-making during problem-solving to learn and develop a good knowledge structure [15, 19, 20, 23, 24, 27, 29-31, 34]. It is important to recognize that engaging in meaningful sense-making to unpack, interpret and apply the laws of physics, which extends and organizes one's knowledge structure and retrieves relevant knowledge to solve future physics problems using similar underlying principles, is an iterative dynamic process (e.g., see Figure 1 for connections). Therefore, appropriate opportunities for metacognition and sense-making while physics problem-solving are necessary to give students an opportunity to refine, repair and extend their knowledge structure and propel them towards a higher level of expertise [15, 19, 20, 23, 24, 27, 29-31, 34]. Metacognitive skills are especially important for learning in knowledge-rich domains such as physics [36-38]. Thus, while solving physics problems[39], opportunities for metacognition, deep sense-making, and knowledge organization (building schema) are only possible [40, 41] if students are provided with appropriate scaffolding support [42, 43] and do not have cognitive overload [44, 45].

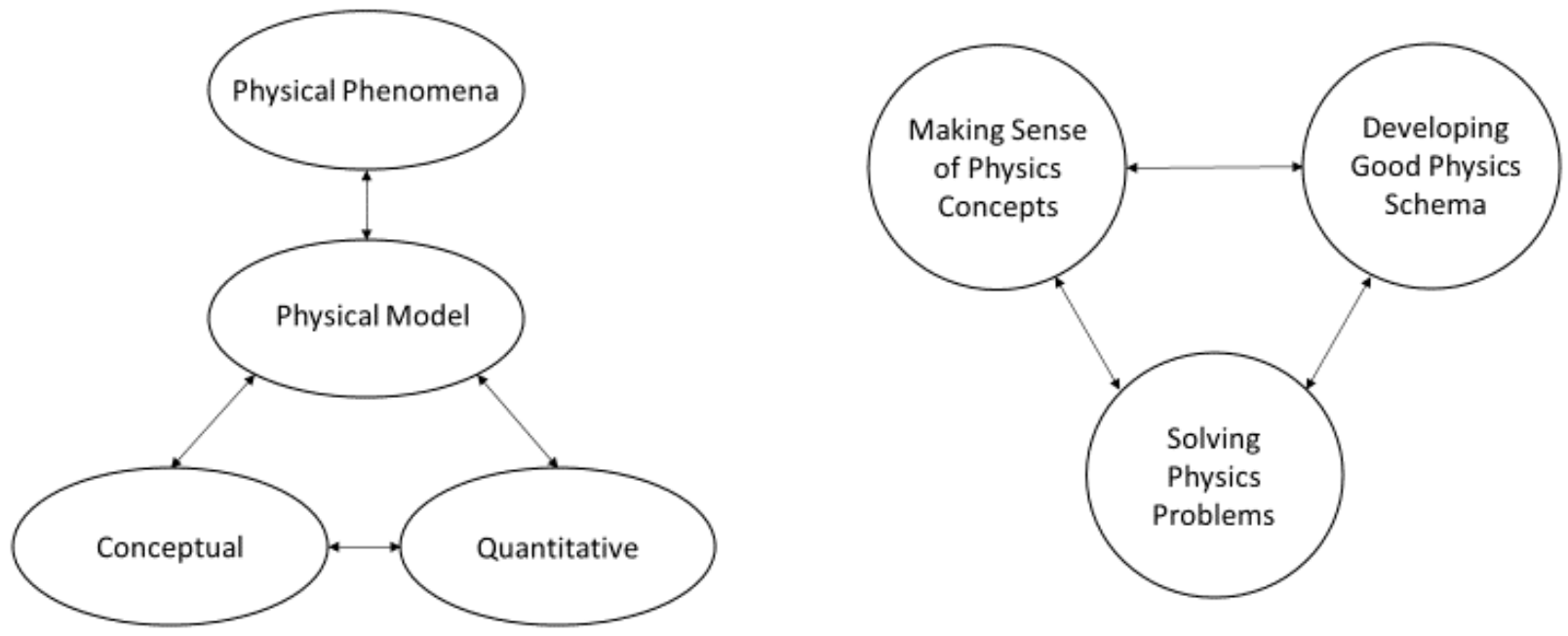

**Figure 1.** Schematic diagrams showing connections between physical phenomena and models which require sense-making with intricate coupling and integration of conceptual and quantitative aspects (**left**), and synergistic components of physics expertise development (**right**).

### *1.2. The ICQUIP Framework and Research Focus*

The "Integrating Conceptual and Quantitative Understanding in Physics" (ICQUIP) framework was developed based on extensive research in physics education showing that expertise requires seamless integration of conceptual and mathematical reasoning. The framework (schematically depicted in Figure 2) emerged from empirical observations that students often treat conceptual and quantitative aspects of physics as separate domains, leading to fragmented understanding. The ICQUIP framework posits that effective physics learning requires deliberate scaffolding to help students connect mathematical formalism with physical meaning, while carefully managing cognitive load to prevent overwhelming students' working memory capacity.

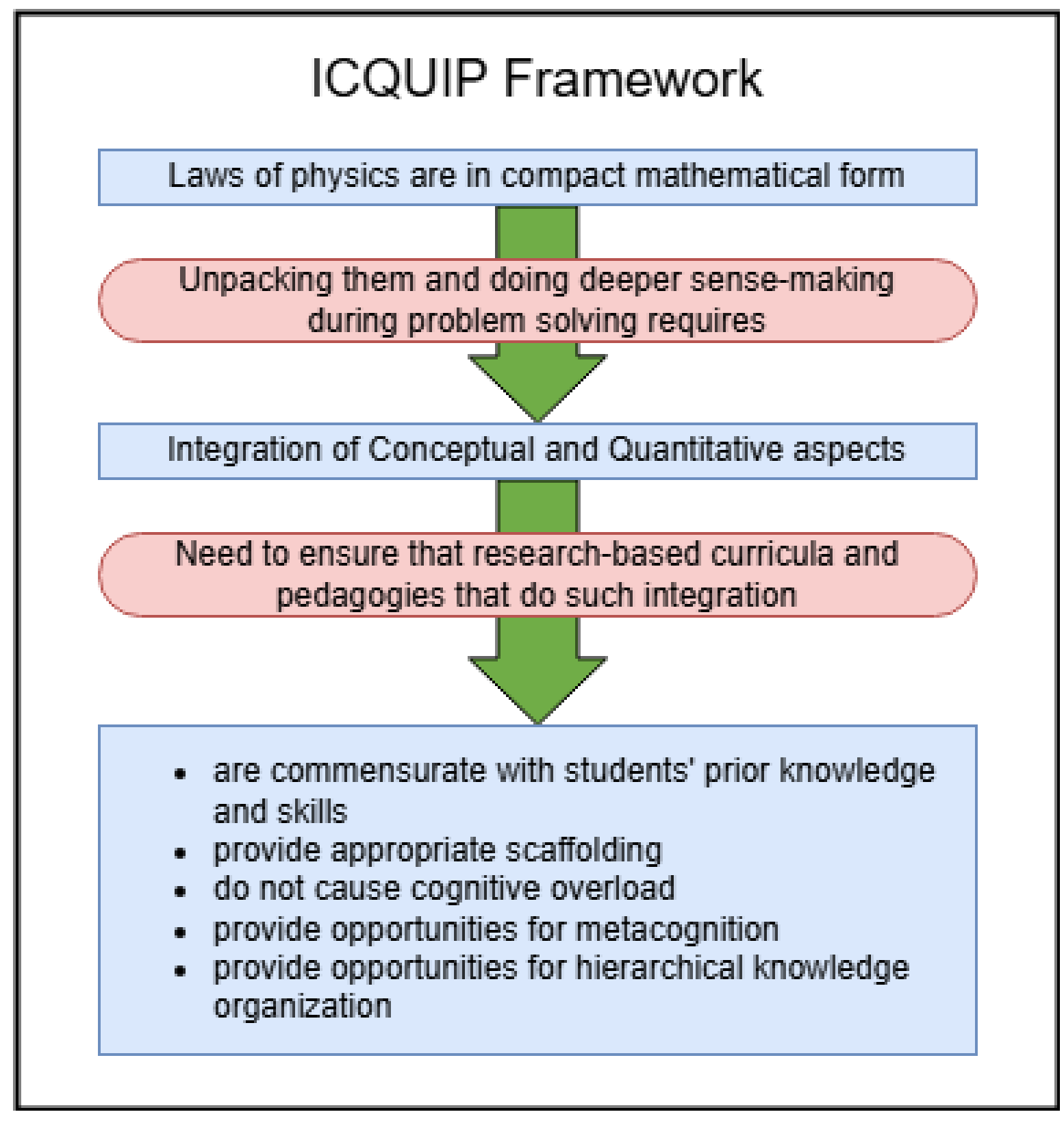

**Figure 2.** Schematic diagram of the ICQUIP framework.

The theoretical foundations of ICQUIP draw from sense-making research focusing on structural integration of physics and mathematics (discussed in the preceding section) as well as cognitive load theory [44], which emphasizes that working memory has limited capacity. Cognitive overload occurs when the intrinsic complexity of the material, combined with the extraneous load from poor instructional design, exceeds students' cognitive capacity. This leaves students with insufficient cognitive resources for germane cognitive load, the mental effort devoted to learning and schema construction [44]. In physics learning, cognitive overload can result from attempting to simultaneously process complex mathematical formalism, physical concepts, and their interconnections without adequate scaffolding support during problem-solving.

Since appropriate conceptual and quantitative connections are central to effective sense-making while problem-solving and developing expertise in physics, we use the ICQUIP framework shown in Figure 2 for the interpretation of results. The framework emphasizes appropriate scaffolding support for the integration of conceptual and quantitative aspects in solving physics problems, commensurate with students' prior knowledge and skills. In particular, the research presented here compares students' expertise measured by their conceptual performance on two versions of a Quantum Interactive Learning Tutorial (QuILT) about quantum optics and investigates the impact of incorporating quantitative reasoning in one version using a framework emphasizing integration of conceptual and quantitative aspects of quantum optics for deeper sense-making [46]. The two versions of the QuILT were developed and validated to help students learn aspects of quantum optics using a Mach Zehnder Interferometer with single photons and polarizers. One version of the QuILT is entirely conceptual while the other version integrates conceptual and quantitative reasoning (hybrid version). We compared the performance on conceptual questions of undergraduate and graduate students utilizing this hybrid QuILT with those utilizing the conceptual QuILT emphasizing the same concepts. Both versions of the QuILT use a scaffolded approach to learning and exploit research on students' difficulties in learning these challenging concepts as a guide. The same conceptual pre-test and post-test were given after traditional lecture-based instruction in relevant concepts and after student engaged with the tutorial, respectively. We note that this paper builds upon a conference proceedings paper [47] which discussed the average performance of the conceptual and hybrid groups but did not discuss the ICQUIP framework or elaborate on the contexts of each of the pre-/post-test questions, how students performed on each of them, and what we can infer about their expertise from it.

The framework is inspired by the fact that since the laws of physics are framed in compact mathematical form, unpacking them and engaging in deeper sense-making would involve intricate coupling of conceptual and quantitative aspects while solving problems [15, 19, 20, 23, 24, 27, 29-31, 34]. However, it is critical to ensure that research-based curricula and pedagogies that do such integration are commensurate with students' prior knowledge and skills, provide appropriate scaffolding support, do not cause cognitive overload, and provide adequate opportunities for metacognition and deeper sense-making as well as opportunities for hierarchical knowledge organization.

The ICQUIP framework stresses that a focus only on conceptual or quantitative aspect may not provide adequate opportunities for sense-making and hierarchical knowledge organization. For example, an excessive explicit focus on quantitative aspects in physics problem-solving can lead many students to memorizing concepts and formulas and solving problems algorithmically using plug-and-chug approaches without deeper sense-making and using problem-solving as an opportunity for knowledge organization. On the other hand, when the focus is only on physics concepts without integration of conceptual and quantitative aspects commensurate with student prior knowledge, students may struggle to engage in deeper sense-making and knowledge organization. The ICQUIP framework's focus on appropriate coupling of conceptual and quantitative aspects for expertise development is particularly important for instruction because physics is a discipline in which quantitative facility and algorithmic approaches can mask a lack of functional understanding and robust knowledge organization. Therefore, to become an expert in physics, students at all levels must be provided with appropriate scaffolding support to integrate conceptual and quantitative aspects of problem-solving in a meaningful way without experiencing cognitive overload. They should be provided support to internalize, for instance, that equations encountered in physics problem-solving are relations between physical quantities that should be used for deeper sense-making and organizing their knowledge rather than treating them as a plug-and-chug tool or a "formula" to obtain an answer. Since many students focus on what they are graded on, a lack of focus on this integration in instructional design and assessment can dis-incentivize engaging in deeper sense-making and making math–physics connections for robust knowledge organization.

Also, conceptual reasoning without the integration of conceptual and quantitative aspects of physics problem-solving can often be more challenging for students who are not physics experts, because an exclusively conceptual problem often lacks levers for scaffolding students' learning, unlike an integrated problem in which quantitative elements can be used as scaffolds to solve the problem by constraint satisfaction before further sense-making and knowledge organization [37, 38]. For example, if a student correctly writes down all equations involved in solving a problem, they can combine them in any order to obtain a quantitative solution before reflecting on their solution so that they can avoid cognitive overload [37, 38]. On the contrary, while reasoning conceptually without quantitative tools, the student must understand the physics underlying the given situation and generally proceed in a particular order to arrive at the correct conclusion [37, 38]. Therefore, the probability of deviating from the correct reasoning chain increases rapidly as the chain of conceptual reasoning becomes long because many students may not have sufficient level of expertise to engage in appropriate conceptual reasoning like physics experts [37, 38]. Combining quantitative and conceptual problem-solving while making the integration commensurate with students' prior knowledge and skills can provide scaffolding for deeper sense-making and knowledge organization. As an example, if asked whether the tension in the cable is greater or smaller than the weight of an elevator accelerating upward, a student who has learned to reason with such integration can invoke Newton's second law in mathematical form *explicitly* to calculate the normal force in terms of the tension and weight. Then, the student can use calculation to reason conceptually and conclude that the tension in the cable is greater than the weight of the elevator accelerating upward. On the other hand, a physics expert can use the same law *implicitly* without explicitly using quantitative reasoning and be confident in conceptually arguing that the upward acceleration implies that tension must exceed the weight. Similarly, if a student does not know whether the maximum safe driving speed while making a turn on a curved horizontal road depends on the mass of the vehicle, they will have difficulty reasoning without using quantitative reasoning that the maximum speed is independent of the mass. However, again in this situation, if students are asked to make conceptual inferences after obtaining a mathematical form for maximum safe driving speed, the scaffolding support from the conceptual and quantitative integration can help them arrive at the correct inference [48], and then additional support can help them do further reflection and sense-making of the entire problem-solving process which is useful for knowledge organization (e.g., why the safe driving speed does not depend on the mass of the vehicle).

Research shows that many students in physics courses who have become facile at quantitative manipulation are unable to solve similar isomorphic problems posed exclusively conceptually [48]. For example, in a study on student understanding of diffraction and interference concepts, the group that was given a quantitative problem performed significantly better than the group given a similar conceptual problem [49]. In another study, Kim et al. examined the relation between traditional physics textbook-style quantitative problem-solving and conceptual reasoning [50]. They found that, although students on average had solved more than 1000 quantitative mechanics problems and were facile at mathematical manipulations, they struggled with conceptual problems on related topics. In another study, when students were given quantitative problems related to power dissipation in a circuit, students performed significantly better than when an equivalent group was given conceptual questions about the relative brightness of light bulbs in similar circuits [51]. In solving the quantitative problems, students applied Kirchhoff's rules to write down a set of equations and then solved the equations algebraically for the relevant variables from which they calculated the power dissipated. When the conceptual circuit question was given to students in similar classes, many students appeared to guess the answer rather than reason about it systematically [51]. Thus, quantitative facility does not imply deep expertise in physics. Prior research [48] also suggests that without scaffolding support, many students are reluctant to convert a problem posed conceptually into a quantitative problem even when explicitly asked to do so. The task of first converting a conceptual problem to a quantitative problem is cognitively demanding [45, 52] even though students were more likely to obtain the correct answer by integrating conceptual and quantitative approaches. The students often used their gut feelings rather than explicitly invoking relevant physics concepts or principles for the conceptual problems. These studies show that in traditional physics courses, students often view quantitative physics problems as "plug-and-chug" exercises while conceptual problems alone are viewed as guessing tasks with little connection to physics concepts [48]. Furthermore, providing students with appropriate scaffolding support to draw conceptual inferences after solving physics problems that explicitly integrate conceptual and quantitative aspects can help them develop expertise, e.g., once students have a quantitative solution, their cognitive resources may be freed up for meaningful sense-making, drawing conceptual inferences, and knowledge organization. The ICQUIP framework's focus on such explicit integration while ensuring that students do not have cognitive overload and have sufficient opportunities for sense-making is important for the design of curricula, because without explicit integration, many students may not automatically harness quantitative problem-solving as an opportunity to reflect upon their solution conceptually and build a good knowledge structure. Prior research shows that problem-solving that combines quantitative and conceptual problems can be an effective instructional strategy consistent with the ICQUIP framework for helping students learn physics [53, 51, 48]. One study showed improvement in students' performance on conceptual problems after solving isomorphic quantitative problems, and many students recognized their similarity and took advantage of their quantitative solution to solve the conceptual problem [48].

Transitioning to learning quantum concepts, in the twenty first century, we are amid the second quantum revolution, and educators at both college and precollege levels are focusing on challenges in teaching quantum concepts and pedagogical approaches to improve students' understanding [54-65]. Prior research suggests that the use of algorithmic approaches for problem-solving in quantum mechanics (QM) courses is common, e.g., see [66, 67]. For example, in surveys administered to 89 advanced undergraduates and more than two hundred graduate students from seven universities enrolled in QM courses, students were given a problem in which the wave function of an electron in a one-dimensional infinite square well of width *a*, at time *t* = 0 in terms of stationary states is given by $\Psi(\mathrm{x}, 0) = \sqrt{\frac{2}{7}}\phi_1 + \sqrt{\frac{5}{7}}\phi_2$Error! Bookmark not defined.. They were asked to write down the possible values of energy and the probability of measuring each value and then calculate the expectation value of energy in the state $\Psi(\mathrm{x}, \mathrm{t})$. Although 67% of the students correctly noted that the probability for the outcome of the ground state energy $E_1$ is $\frac{2}{7}$ and the first excited state energy $E_2$ is $\frac{5}{7}$, the majority were unable to use this information to determine the expectation value of energy $\frac{1}{7}(2E_1 + 5E_2)$. Not only did the correct responses decrease from 67% to 39%, but the students who calculated the expectation value $\langle E \rangle$ correctly also mainly exploited brute-force methods: first writing $\langle E \rangle = \int_{-\infty}^{\infty} \langle \Psi | \hat{H} | \Psi \rangle \, dx$, then expressing $\Psi(\mathrm{x}, \mathrm{t})$ in terms of the two energy eigenstates, acting $\hat{H}$ on the eigenstates, and finally using orthogonality to obtain the final answer. Some were lost early in this process while others did not remember some other mechanical steps, e.g., taking the complex conjugate of the wave function, using orthogonality of stationary states, or not realizing the proper limits of the integral. Interviews revealed that some students did not know or recall the conceptual interpretation of the expectation value as an ensemble average and did not realize that the expectation value of energy could be calculated more simply in this case by taking advantage of the first part. Other problems posed to advanced students in QM courses confirmed that many of the difficulties are conceptual in nature [66, 67]. Analogous difficulties were also observed in response to conceptual problems about Larmor spin precession, e.g., regarding the expectation values of spin components and their time dependence, given a particular initial state [68]. These examples demonstrate that quantitative facility does not automatically imply conceptual understanding in QM, and students can perform well on quantitative problems using algorithmic approaches without understanding the underlying QM concepts. In this sense, from the introductory to advanced levels, physics knowledge organization and conceptual understanding, critical for physics expertise development, are challenging for students even if they show facility with algorithmic problem-solving. Furthermore, like the algorithmic approaches to problem-solving in introductory physics, strict quantitative exercises in advanced courses like QM often fail to provide adequate incentives and opportunities for sense-making and drawing meaningful inferences from the problem-solving process to repair, organize, and extend one's knowledge.

One effective approach for developing and implementing research-based curricula that integrate conceptual and quantitative problem-solving consistent with the ICQUIP framework is to provide students scaffolding support to engage in explicit quantitative aspects of problem-solving followed by more conceptual aspects of problem-solving (that ask students to draw conceptual inferences) based upon a pedagogical analysis of knowledge inherent in quantitative problems to avoid cognitive overload. Particular attention should be given to designing these integrated problems such that the conceptual aspects of the problems challenge students to differentiate between related concepts which can easily be misinterpreted. The scaffolding provided while engaging in sense-making can focus on strategies for making conceptual inferences from symbolic expressions and helping students contemplate on how they can use them to extend/repair their knowledge structure[37]. Scaffolding support can be a critical component of the design of these types of research-based curricula to ensure that students do not have cognitive overload and develop independence, e.g., initially they can be guided frequently with prompt feedback and support, but the frequency of scaffolding support can decrease as students become better at making conceptual inferences. These considerations were central to the hybrid Mach Zehnder Interferometer QuILT discussed here.

### *1.3. Developing Expertise in Quantum Optics Using Mach Zehnder Interferometer and Goal of the Investigation*

While learning physics is challenging even at the introductory level because it requires drawing meaningful inferences and unpacking and applying the few physics principles, which are in compact mathematical forms, to diverse situations[37], learning advanced physics is also challenging because one must continue to build on all the prior knowledge acquired at the introductory and intermediate levels. In addition, the mathematical sophistication required is generally higher for advanced physics. To learn, students must focus on the physics concepts while solving problems and must be able to go back and forth between mathematics and physics seamlessly, regardless of whether they are converting a physical situation to a mathematical representation or contemplating the physical significance of a quantitative solution.

Furthermore, research suggests that learning QM is especially non-intuitive and challenging for advanced students. Prior research also demonstrates that the patterns of difficulties in the context of QM bear a striking resemblance to those found in introductory classical mechanics [69]. These analogous patterns of difficulties are often due to the “paradigm shift” from classical mechanics to QM as well as diversity, e.g., in prior preparation and goals of upper-level students [69]. For example, unlike classical physics, in which position and momentum are deterministic variables, in QM they are operators that act on state vectors, which lie in an abstract Hilbert space. The different paradigms of classical mechanics and QM suggest that even students with a good knowledge of classical mechanics will start as novices and gradually build their knowledge structure about QM with scaffolding support.

We chose the Mach Zehnder Interferometer (MZI) with single photons and polarizers as the context for this investigation. One reason the MZI with single photons was selected is that in recent decades, quantum optics has become a thriving area of research, with single-photon experiments playing a key role in the quantum information revolution, and educators are considering pedagogical approaches to help students learn relevant concepts [70-82, 47, 83-94]. The MZI experiments with single photons and polarizers provide students with an opportunity to learn the fundamentals in a concrete situation (e.g., see ref. [95]), e.g., these experiments elegantly illustrate the single-photon wave–particle duality, interference of single photons, and the fact that quantum measurements are inherently probabilistic. However, these concepts are challenging even for advanced students.

The investigation reported here is based on the hypothesis that, consistent with the ICQUIP framework, integrating conceptual and quantitative aspects in a research-based curriculum can help students build a more coherent knowledge structure of QM if students are provided with appropriate scaffolding support, do not have cognitive overload, and have adequate opportunities for sense-making. The goal of the research is to analyze the impact of incorporating quantitative reasoning in a Quantum Interactive Learning Tutorial (QuILT), which consists of research-validated inquiry-based learning sequences, on students' conceptual understanding of quantum optics in the context of the Mach Zehnder Interferometer (MZI) with single photons and polarizers [96, 95, 97], and to interpret the findings based on the ICQUIP framework. We developed and validated two versions of the QuILT on the MZI with single photons and polarizers that strive to help students learn about foundational issues in quantum optics. One version only requires conceptual reasoning while the second version uses an integrated conceptual and quantitative approach (called the conceptual and hybrid versions, respectively, for convenience). The conceptual QuILT focuses on engaging students with conceptual reasoning only under the assumption that the underlying quantum mechanics concepts involving single-photon interference and quantum eraser may be sufficiently complex so that incorporating quantitative aspects involving product states of path and polarization may cause cognitive overload for students [52, 98]. Apart from PERC and GIREP conference proceedings, e.g., in ([97]) which used the MZI as a context to elaborate with examples on the iterative process of development and validation of a QuILT in general, we previously described the findings for the conceptual QuILT with a somewhat greater focus on the underlying physics for educators [96] and on the framework and "warmup" of the QuILT for researchers [95] to reach different stakeholders [99].

This paper builds on a conference proceedings [47] and uses ICQUIP framework to compare the conceptual QuILT with the hybrid QuILT. The ICQUIP framework posits that it is difficult to help students develop a good knowledge structure of physics without integrating conceptual and quantitative aspects in research-based tools because quantitative tools can provide constraints to help students engage in deeper sense-making so long as they do not have cognitive overload and can provide students with opportunities for metacognition[37]. The conceptual and hybrid versions of the QuILT both use a scaffolded inquiry-based approach to learning, but the hybrid QuILT focuses on helping students make conceptual inferences after obtaining quantitative solutions. The learning objectives of the hybrid QuILT pertaining to the level of conceptual understanding desired are the same as those for the conceptual QuILT[95]. Moreover, before students engaged with either the conceptual or the hybrid QuILT, we investigated their conceptual understanding of quantum optics after traditional lecture-based instruction (pre-test). After traditional instruction in relevant concepts and after the conceptual/hybrid QuILT, students were administered the same conceptual pre-/post-tests (see Appendix A) to evaluate their conceptual understanding. Students engaged with the conceptual or hybrid QuILT on quantum optics in consecutive years. Our goal was to compare the extent to which students who engaged with the conceptual and hybrid versions of the QuILT improved in their performance on the underlying concepts compared to traditional instruction and to interpret the data based on the ICQUIP framework.

We stress that the interpretation of the MZI experiments using quantitative tools in the hybrid QuILT involves making conceptual inferences using product states of path and polarization, although the underlying concepts can be taught using only conceptual reasoning as in the conceptual version of the QuILT[95]. We also note that both the conceptual and hybrid versions of the MZI QuILT encourage students to use visualization tools (computer simulations developed by Albert Huber) using an MZI including the counter-intuitive quantum eraser experiment. Both versions of the QuILT focus on using the same contexts involving the MZI experiment to help students learn concepts such as the wave–particle duality of a single photon, interference of a single photon with itself, and the probabilistic nature of quantum measurements. Students also learn how adding photodetectors and optical elements such as beamsplitters and polarizers in the paths of the MZI affect the measurement outcomes. Supplementary Materials describe the background of MZI experiments with and without polarizers as well as the specific MZI experimental contexts students engage with in both versions of the QuILT (see Figure S2 in the Supplementary Materials) and any differences between the two versions. The main difference between the two versions is that in the hybrid version, some of the guided learning sequences involving conceptual reasoning *only* are replaced by those sequences involving both quantitative and conceptual reasoning. This is performed with the assumption that advanced students in quantum mechanics courses will benefit from the opportunity to make conceptual sense-making from their quantitative solutions and will not have a cognitive overload despite the increased mathematical sophistication of the hybrid QuILT because the learning sequences are research-validated and appropriately scaffolded. We interpret the results using the ICQUIP framework in Section 3.

Figure 3 shows two arrangements from pre-/post-tests (see Appendix A for all questions) that share a surface feature with the quantum eraser [95]. However, 45° polarized single photons emitted from the source and propagating through either of these arrangements do not display interference in the way that the quantum eraser does because of which-path information (WPI) for all photons. We note that in both versions of the QuILT, students engaged in a situation like that in Figure 3 on the left-hand side (with a vertical polarizer right before detector D1). However, the situation in the right-hand side of Figure 3, which was posed as a question on the pre-/post-tests, was novel for all students in that they had not encountered this situation in traditional instruction or in the QuILT. We wanted to use this context to investigate transfer of learning from the situations students had learned to a new situation [41, 100-102].

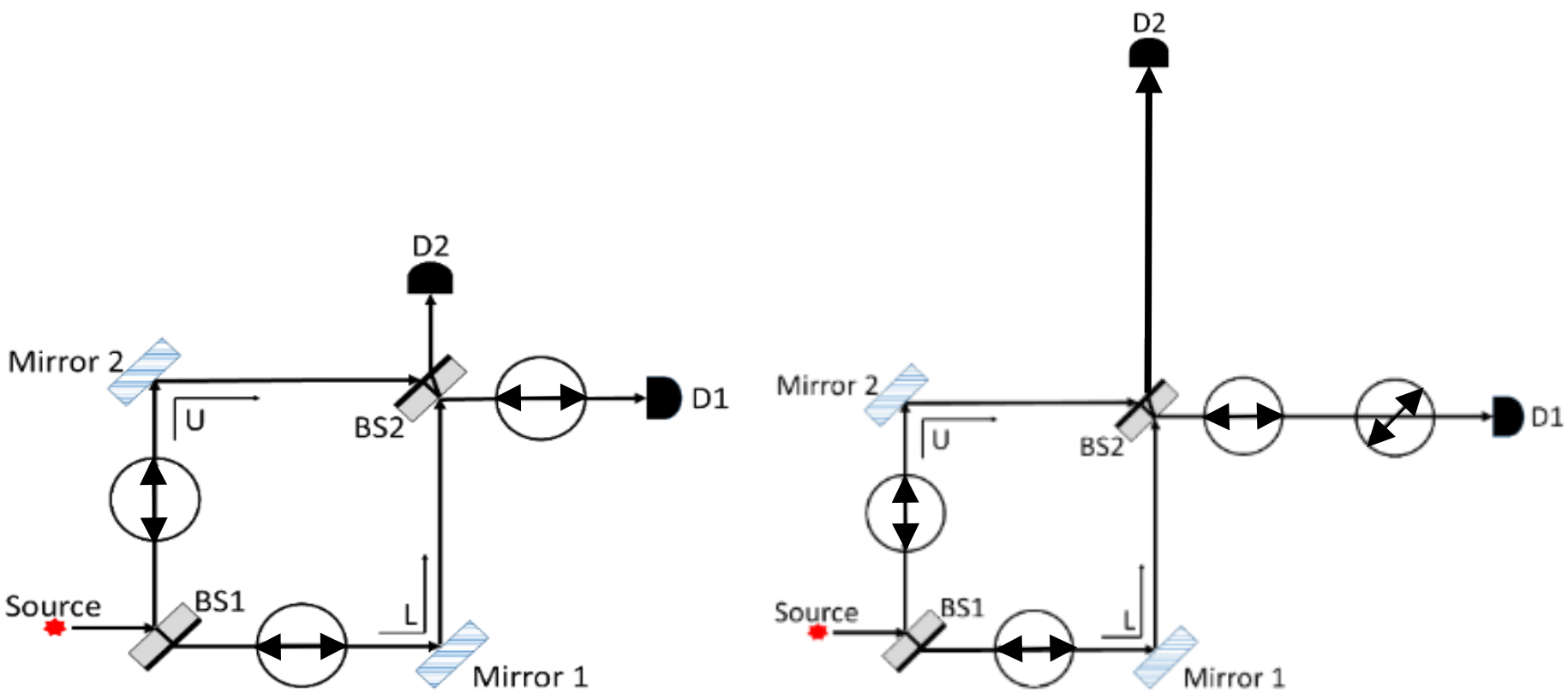

**Figure 3.** The MZI arrangement similar to the quantum eraser set up[95], but with a horizontal polarizer in place of the 45° polarizer between BS2 and D1 (**left**) and an additional horizontal polarizer in addition to the 45° polarizer between BS2 and D1 as shown (**right**).

Before we describe the organization of the rest of the paper, we note that the Supplementary Materials not only provide a brief background on the MZI experiment but also briefly summarize the development and validation of the two versions of the QuILT and how they strive to address students' difficulties (these are analogous for both versions, and the methodology for the conceptual version is described in Ref. [95]). Below, we discuss the research objective, research questions, and our hypotheses, and then the methodology including the in-class implementation of the QuILT and comparison of the student pre-test (after traditional lecture-based instruction) and post-test (after engaging with the QuILT) average scores for the two versions of the QuILT. Then, we present our Results and Discussion Section with the finding showing that graduate students typically performed at least as well or better on conceptual problems on post-test after engaging with the hybrid MZI QuILT that combines conceptual and quantitative problem-solving compared to when they engaged with the conceptual-only QuILT but the results for undergraduates differ depending on their pre-test performance. This is followed by a broader Discussion Section focusing on the interpretation of the results using the ICQUIP framework, limitations and future directions and then a Conclusion Section.

### *1.4. Research Objective, Questions and Hypotheses*

As we mentioned in the preceding sections, the primary objective of this study is to investigate the impact of integrating conceptual and quantitative reasoning in a research-based learning tool (QuILT) on students' conceptual understanding of quantum optics, using the ICQUIP framework as a theoretical lens. Specifically, we address the following research questions:

**RQ1.** *How does the integration of conceptual and quantitative reasoning in a hybrid QuILT affect graduate students' conceptual understanding of quantum optics compared to a purely conceptual QuILT?*

**RQ2.** *How does the integration of conceptual and quantitative reasoning in a hybrid QuILT affect undergraduate students' conceptual understanding of quantum optics compared to a purely conceptual QuILT?*

**RQ3.** *What role does students' prior knowledge play in determining whether they benefit from the integrated conceptual–quantitative approach versus the conceptual-only approach?*

Based on the ICQUIP framework, we hypothesize the following:

**H1.** *Graduate students will benefit more from the hybrid QuILT than the conceptual QuILT due to their stronger mathematical facility, which can reduce cognitive load and allow for deeper conceptual processing.*

**H2.** *Undergraduate students' benefit from the hybrid QuILT will depend on their prior conceptual understanding, e.g., those with stronger preparation will benefit more from integration, while those with weaker preparation may experience cognitive overload since the scaffolding provided may be inadequate for them.*

**H3.** *The effectiveness of integrating conceptual and quantitative reasoning will be most apparent for complex problems requiring an understanding of four-dimensional Hilbert space (involving both path and polarization states).*

## 2. Methodology

### *2.1. Participants*

The study involved physics students from a large research university in the United States across multiple years. In particular, the participants of this study were students in a mandatory junior–senior level quantum course or physics graduate level course**.** Regarding undergraduate students' mathematical preparation, students would have taken calculus I, II, and III (vector calculus), differential equation, and linear algebra courses with at least a C grade (2 points on a 4-point scale) before this QM course since those are prerequisites for the QM course. All instructors were supportive of physics education research and had implemented physics education research-based pedagogies in their classes.

The graduate student sample consisted of the following:

- Hybrid QuILT group: N = 10 (matched pre/post), first-year physics Ph.D. students
- Conceptual QuILT group: N = 27 (matched pre/post), first-year physics Ph.D. students enrolled simultaneously in a graduate QM course and Teaching of Physics course
- Note: Conceptual QuILT group in Ref. [96, 95]: N = 45 (matched pre/post), first-year physics Ph.D. students enrolled simultaneously in a graduate QM course and Teaching of Physics course (these results for two years administration only include a subset of pre-/post-questions)

The undergraduate student sample consisted of the following:

- Hybrid QuILT group A: N = 24 (pre-test), N = 20 (post-test), junior/senior physics majors
- Hybrid QuILT group B: N = 15 (pre-test), N = 16 (post-test), junior/senior physics majors
- Conceptual QuILT group: N = 26 (pre-test), N = 20 (post-test), junior/senior physics majors

Note: Conceptual QuILT group in Ref. [96, 95]: N = 44 (pre-test), N = 38 (post-test), junior/senior physics majors (these results for two years administration only include a subset of pre-/post-questions).

We did not explicitly ask for gender and age information from students, but based upon the demographics at the university, there are 20–25% women in both graduate and undergraduate physics courses. Since undergraduate students were juniors and seniors, a majority of them are likely to be 21–22 years old. The graduate students were in their first year of Ph.D. program, so most are likely to be older than 22 years old; most are typically in their 20s.

### *2.2. Materials*

#### 2.2.1. QuILT Versions

Two versions of the Mach–Zehnder Interferometer QuILT were developed through iterative design and validation involving both expert and student feedback.

Conceptual QuILT: Focuses exclusively on conceptual reasoning about single-photon interference, which-path information, and measurement outcomes in various situations involving MZI without mathematical formalism.

Hybrid QuILT: Integrates conceptual reasoning with quantitative tools including 2 × 2 matrix mechanics for path states and 4 × 4 matrix mechanics for product states for combined path-polarization cases. The hybrid version scaffolds students to make conceptual inferences from mathematical solutions.

The learning objectives of both versions of the QuILT regarding the concepts we want students to learn are the same and reflected in the pre-test/post-test questions in Appendix A. The details of the conceptual version are discussed in Ref.[95]. Both the conceptual and hybrid versions cover the same concepts except that in the hybrid QuILT, the case with a photodetector in one of the two paths of the MZI was not covered. Both versions of the QuILT can be found at PhysPort[103]. The hybrid QuILT begins with a warm-up that builds on students' prior knowledge about the interference of light and then helps students learn about the MZI with single photons using an integrated conceptual and quantitative approach requiring only 2 by 2 matrix mechanics with the upper and lower path states. Then, students transition to the main section of the QuILT that focuses on the fundamentals of quantum mechanics in the context of the MZI with single photons and polarizers using an integrated conceptual and quantitative approach, requiring conceptual interpretations of the product states of path and polarization. Students are provided with scaffolding support as they construct matrices that describe different elements of the MZI. Students then use these matrices to describe various arrangements in each physical situation and engage in sense-making of the underlying concepts using quantitative tools. Using an integrated conceptual and quantitative approach throughout the QuILT, students make predictions about a particular MZI set up, and work through integrated conceptual and quantitative learning sequences. At the beginning of the QuILT, students are asked to check their predictions via a computer simulation available and reconcile the differences between their predictions and observations. Students are also given opportunities to reflect on the concepts learned throughout to ensure sense-making and opportunities for knowledge organization.

#### 2.2.2. Assessment Instrument

The assessment instrument used for pre-/post-test was the same and validated through expert review and student interviews. The pre-/post-test consisted of 11 open-ended conceptual questions about MZI experiments (see Appendix A). The conceptual problems on both the pre-test and post-test, which are identical, are open-ended (see Appendix A). The open-ended format requires that students generate answers based on an understanding of the concepts as opposed to memorization of the concepts. Questions assessed their understanding of the following concepts:

Single-photon interference (Q1–Q3: 2D Hilbert space);

Role of detector in one of the paths (Q4);

Effects of polarizers on the interference at the detector (Q5–Q10: 4D Hilbert space);

Experimental design related to MZI (Q11).

The hybrid QuILT includes the same MZI experiments as the conceptual QuILT ([96, 95]) (see Supplementary Materials Figure S2 and discussion) except that a situation like that shown in pre-/post-test Q4 in Appendix A with a detector in the U path was not in the hybrid QuILT, but there was an isomorphic situation that students engaged with in the conceptual version of the QuILT with a detector in one of the paths (Ref. [96]).

The same rubric as in Ref. ([96, 95]) was used to grade the pre-test and post-test of students who used either the conceptual or the hybrid version of the QuILT. Two graders graded the pre-/post-tests on a jointly developed rubric, and the inter-rater agreement was above 95%.

### *2.3. Procedure*

The study followed the following timeline. The information regarding the order in which the different components of the two forms of the QuILT and the pre-/post-test were administered is summarized in Figure 4, and additional information about the particular MZI experiments that students engaged with in both the hybrid and conceptual versions of the QuILT is in Figure S2 in the Supplemental Materials. Thus, both versions of the MZI QuILT with single photons include the same pre-test which was given to students after lecture-based instruction on the relevant concepts but before students worked on the QuILT, while the post-test was given after students completed the QuILT. A single facilitator conducted all in-class QuILT sessions to ensure consistency.

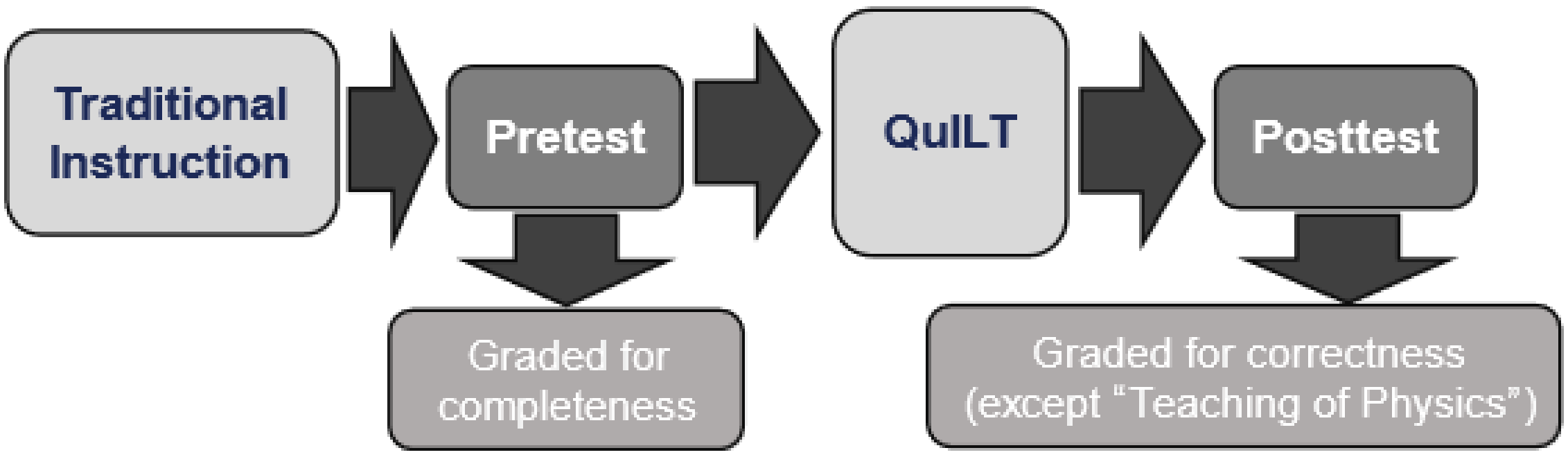

**Figure 4.** A summary of the order of in-class implementation.

Traditional instruction: Instructors covered MZI concepts through lectures within one week.

Pre-test: Pre-test was administered after traditional instruction to assess baseline understanding (in class, 50 min). The goal of the pre-test is to assess student baseline knowledge after traditional lecture-based instruction.

QuILT implementation: Students completed the warm-up section as homework, then the main QuILT was partially completed in class with a facilitator present before the students completed the remainder as homework.

Post-test: Post-test was administered in class in a 50 min period soon after QuILT completion (within 1–2 weeks), depending on the instructor's other constraints related to the course.

As summarized earlier, the hybrid QuILT was implemented in two upper-level undergraduate QM courses (spanning two consecutive years) and one first-year graduate core quantum mechanics course, which first received traditional lecture-based instruction on the relevant concepts. The instructors for all these courses were different. The course instructors provided an overview of the MZI set up, covering key concepts such as the propagation of light through beamsplitters, the phase differences introduced by the two paths in the MZI, the interpretation of detector clicks, and the effects of polarizers placed at various locations within the MZI. These foundational topics aligned with the experiments students would encounter in the QuILT. We note that we did not have control over how each instructor would lecture on these concepts prior to the pre-test, although they were advised on the essential content to cover before administering the pre-test and were provided with recommendations for incentivizing student participation in the pre-test, QuILT, and post-test. The pre-test was administered during class, followed by the QuILT warm-up, which students completed as homework. Students then worked through part of the hybrid QuILT in class, with one week allocated to complete the remaining sections as homework. The in-class portion of the QuILT was facilitated across all courses by one of the authors of this article, who acted as a guest instructor. This facilitator had also conducted the in-class conceptual MZI QuILT sessions in Ref. ([96, 95]), which provides some consistency in implementation. For both undergraduate and graduate students, the pre-test and QuILT constituted a small portion of their overall homework grade. After completing the QuILT, students took the post-test during their respective QM classes. Ample time was provided for all students to complete both the pre-test and post-test. The post-tests for the hybrid QuILT were graded for correctness as a quiz for both the undergraduate and graduate QM courses.

We note that the data from undergraduates in Ref. ([96, 95]) are for the conceptual MZI QuILT implemented in the same course for physics juniors and seniors at the same university as the hybrid QuILT implementation but in a different year. The graduate student data in Ref. ([96, 95]) were collected from first-year graduate students enrolled simultaneously in a graduate core QM course and a semester-long Teaching of Physics course, which was a mandatory pass/fail course. The conceptual QuILT pre-test and post-test were incorporated into the Teaching of Physics course to introduce graduate students to the tutorial-based approach to learning and teaching. This approach was adopted because the instructor of the core QM course declined to administer these activities due to time limitations. Thus, although the conceptual MZI QuILT was administered in a similar manner to the hybrid MZI QuILT to a similar graduate student population in different years at the same university (first semester physics graduate students), since the conceptual QuILT was administered in a pass/fail course, pre-/post-tests and the conceptual MZI QuILT were graded for completeness ([96, 95]). We note that the data that were previously reported in Refs. ([96, 95]) are for two consecutive years of the course for some of the test items in Appendix A. In addition, here we also compare students' performance after engaging with the two versions of the QuILT for some test items that are not in Refs. ([96, 95]) and are only for one year of administration for graduate and undergraduate students in the conceptual QuILT group.

### 2.4. How the QuILT Supports Metacognition

Both QuILT versions incorporate features designed to promote metacognitive awareness and self-regulated learning. They have prediction–scaffolding/observation–reconciliation cycles such that students predict outcomes, then scaffolding is provided (in some cases it could be students engaging with the simulation provided), and then they are asked to reconcile discrepancies. They have reflection prompts asking students questions to explain their reasoning and identify conceptual changes. They have self-monitoring checkpoints which provide regular opportunities to assess understanding before proceeding. Furthermore, they have scaffolded complexity in that problem difficulty is gradually increased to allow students to monitor their readiness for more complex scenarios. The hybrid QuILT additionally promotes metacognition through explicit prompts to connect mathematical results to physical meaning and opportunities to check mathematical solutions against conceptual predictions and reflection.

### 2.5. Data Analysis

After the in-class implementation, we conducted a comparative analysis of the conceptual and hybrid QuILT data. Each question sub-part was scored 0–100% based on conceptual correctness. Question scores were averaged across sub-parts. For statistical measures, we used normalized gain and effect size which complement each other. For each case, the normalized gain [104] is defined as g = (post-test% − pre-test%)/(100% − pre-test%). For each case, the effect size given by Cohen's $d$ [105] is defined as $d$ = (M_post − M_pre)/SD_pooled, where M_post and M_pre are the average post-test and pre-test scores and SD_pooled is the pooled standard deviation. Due to small sample sizes ($N < 30$ per group), we report descriptive statistics and effect sizes rather than use inferential tests. Our comparisons focus on practical significance (from an instructional point of view: what would be considered meaningful differences by instructors) through effect sizes and normalized gains rather than statistical significance.

## 3. Results and Discussion

Table 1 presents average percentage scores of each group of students on the pre-/post-tests before and after engaging with the hybrid QuILT. The normalized gains and effect sizes (Cohen's $d$) are shown for each class for each conceptual problem. The results from two years of administration of the conceptual QuILT for many of the problems are in Refs. [96]and reproduced in Table A1 in Appendix A. Table 2 shows average percentage scores for additional problems not in Refs. [96, 95] for one year of administration of the conceptual QuILT. Figure 5 shows bar graphs of the pre-test and post-test performance for the three groups that were engaged with the hybrid QuILT (a and b) and with the conceptual QuILT (c and d) for the second year of administration when we had data on all problems. Figures 6 and 7 provide a visual comparison of results in Table 1 with Table A1 in Appendix A from Refs. [96, 95] and Table 2. Below, we discuss the results in the order of our research questions. The performance of all groups, including the ones that engaged with the conceptual QuILT on individual questions, is discussed for deeper insight.

**Table 1.** Percentages on pre-test/post-test questions before and after the hybrid QuILT, normalized gains, and effect sizes (Cohen's *d*). Graduate students are matched (since there was a large difference in N between pre and post, the averages for matched and unmatched are different), while undergraduates had only small fluctuations in participation in pre-/post-test, and there were no significant differences between matched and unmatched results, so all undergraduates are kept in data analysis.

| | **Graduate Students** | | | | **Undergraduates Group A** | | | | **Undergraduates Group B** | | | |
|---|---|---|---|---|---|---|---|---|---|---|---|---|
| **Q** | **Pre(%) N = 10** | **Post(%) N = 10** | **<g>** | ***d*** | **Pre(%) N = 24** | **Post(%) N = 20** | **<g>** | ***d*** | **Pre(%) N = 15** | **Post(%) N = 16** | **<g>** | ***d*** |
| 1 | 55 | 100 | 1.00 | 0.67 | 17 | 93 | 0.91 | 1.21 | 48 | 91 | 0.81 | 0.54 |
| 2 | 55 | 82 | 0.60 | 0.40 | 69 | 89 | 0.64 | 0.40 | 57 | 84 | 0.64 | 0.46 |
| 3a | 60 | 95 | 0.89 | 0.49 | 34 | 100 | 1.00 | 1.18 | 55 | 100 | 1.00 | 0.74 |
| 3b | -- | -- | -- | -- | -- | -- | -- | -- | 28 | 94 | 0.92 | 0.53 |
| 3 (avg) | 60 | 95 | 0.89 | 0.49 | 34 | 100 | 1.00 | 1.18 | 42 | 97 | 0.95 | 1.89 |
| 4a | 10 | 82 | 0.80 | 2.33 | 44 | 60 | 0.29 | 0.34 | 77 | 100 | 1.00 | 0.92 |
| 4b | 20 | 90 | 0.88 | 1.98 | 29 | 56 | 0.36 | 0.54 | 73 | 100 | 1.00 | 0.85 |
| 4c | 30 | 82 | 0.74 | 1.22 | 13 | 56 | 0.49 | 1.01 | 40 | 94 | 0.90 | 1.39 |
| 4 (avg) | 20 | 85 | 0.81 | 2.06 | 29 | 57 | 0.39 | 0.70 | 63 | 98 | 0.95 | 1.36 |
| 5 | 64 | 80 | 0.45 | 0.37 | 56 | 75 | 0.43 | 0.49 | 73 | 97 | 0.88 | 0.73 |
| 6a | 5 | 68 | 0.67 | 3.21 | 10 | 60 | 0.55 | 1.97 | 23 | 47 | 0.31 | 0.69 |
| 6b | 36 | 85 | 0.76 | 1.19 | 48 | 95 | 0.90 | 1.29 | 80 | 81 | 0.06 | 0.03 |
| 6c | 36 | 73 | 0.57 | 0.83 | 23 | 65 | 0.55 | 1.20 | 40 | 59 | 0.32 | 0.50 |
| 6 (avg) | 26 | 75 | 0.66 | 1.83 | 27 | 73 | 0.63 | 1.97 | 48 | 62 | 0.27 | 0.53 |
| 7a | 5 | 95 | 0.95 | 6.32 | 40 | 83 | 0.71 | 1.09 | 77 | 91 | 0.60 | 0.44 |
| 7b | 18 | 90 | 0.88 | 2.07 | 13 | 40 | 0.31 | 0.66 | 40 | 88 | 0.79 | 1.13 |
| 7c | 41 | 91 | 0.85 | 1.29 | 13 | 35 | 0.26 | 0.55 | 20 | 81 | 0.77 | 1.55 |
| 7 (avg) | 21 | 92 | 0.90 | 2.85 | 22 | 53 | 0.39 | 0.93 | 46 | 87 | 0.75 | 1.36 |
| 8a | 23 | 91 | 0.88 | 2.21 | 17 | 65 | 0.58 | 1.53 | 47 | 88 | 0.77 | 1.44 |
| 8b | 18 | 90 | 0.88 | 2.07 | 4 | 80 | 0.79 | 2.40 | 53 | 97 | 0.93 | 1.20 |
| 8c | 9 | 82 | 0.80 | 2.13 | 2 | 58 | 0.57 | 1.79 | 23 | 88 | 0.84 | 1.99 |
| 8 (avg) | 17 | 88 | 0.86 | 2.62 | 8 | 68 | 0.65 | 2.39 | 41 | 91 | 0.85 | 2.16 |
| 9a | 18 | 91 | 0.89 | 2.39 | 15 | 68 | 0.62 | 1.59 | 40 | 94 | 0.90 | 1.86 |
| 9b | 27 | 90 | 0.86 | 1.65 | 8 | 45 | 0.40 | 0.91 | 30 | 88 | 0.82 | 1.48 |
| 9c | 27 | 91 | 0.88 | 1.70 | 4 | 50 | 0.48 | 1.20 | 27 | 84 | 0.79 | 1.46 |
| 9 (avg) | 24 | 91 | 0.88 | 2.19 | 9 | 54 | 0.50 | 1.44 | 32 | 89 | 0.83 | 2.04 |
| 10a | 14 | 91 | 0.89 | 2.59 | 15 | 45 | 0.36 | 0.77 | 30 | 84 | 0.78 | 1.66 |
| 10b | 27 | 90 | 0.86 | 1.65 | 4 | 35 | 0.32 | 0.84 | 33 | 75 | 0.63 | 0.92 |
| 10c | 27 | 82 | 0.75 | 1.31 | 4 | 35 | 0.32 | 0.84 | 33 | 69 | 0.53 | 0.76 |
| 10 (avg) | 23 | 88 | 0.84 | 2.36 | 8 | 38 | 0.33 | 1.05 | 32 | 76 | 0.64 | 1.34 |
| 11 | 18 | 73 | 0.67 | 1.31 | 4 | 70 | 0.69 | 2.08 | 36 | 94 | 0.90 | 1.54 |
| Average | 33 | 87 | 0.80 | 3.46 | 27 | 70 | 0.60 | 2.61 | 48 | 88 | 0.77 | 2.46 |

**Table 2.** Percentages on pre-test/post-test after averaging over the sub-parts of questions that are not in Ref. [96] before and after the conceptual QuILT, normalized gains, and effect sizes. Results are for 27 graduate students (matched) and 26 undergraduates in pre-test and 20 in post-test who engaged with the conceptual QuILT. Results discussed in [96] are provided in Appendix A.

| | Graduate Students | | | | Undergraduates | | | |
|---|---|---|---|---|---|---|---|---|
| Q | Pre(%) | Post(%) | <g> | *d* | Pre(%) | Post(%) | <g> | *d* |
| 5 | 9 | 59 | 0.55 | 1.26 | 31 | 85 | 0.78 | 1.31 |
| 6 | 30 | 77 | 0.67 | 1.74 | 34 | 89 | 0.84 | 2.16 |
| 9 | 35 | 74 | 0.60 | 1.15 | 11 | 83 | 0.80 | 2.63 |
| 10 | 35 | 53 | 0.28 | 0.48 | 12 | 59 | 0.54 | 1.43 |
| 11 | 26 | 48 | 0.30 | 0.48 | 8 | 70 | 0.68 | 1.66 |

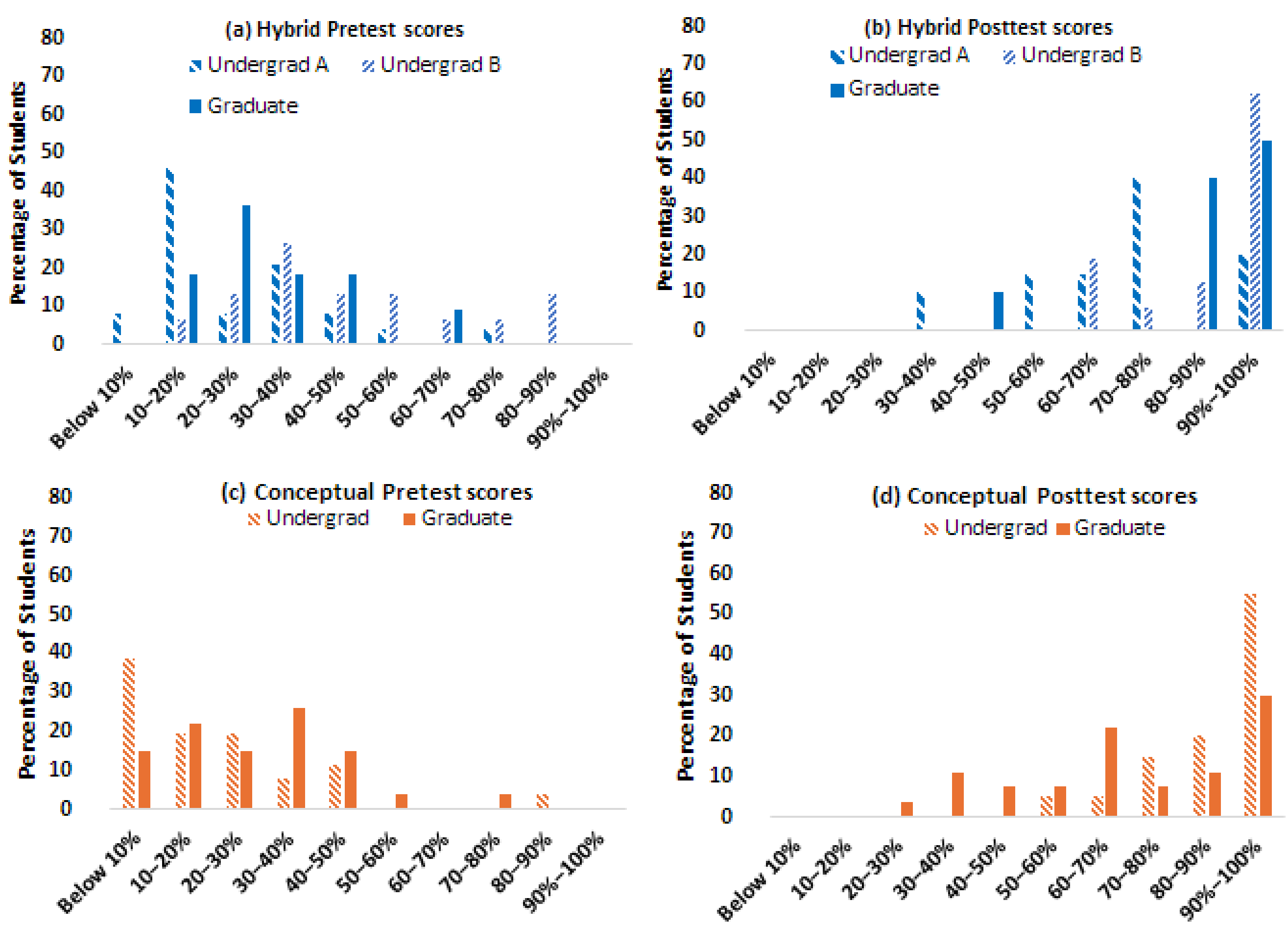

**Figure 5.** Total score distributions (**a**) (top left) pre-test; (**b**) (top right) post-test for each hybrid QuILT group [47]. Total score distributions (**c**) (bottom left) pre-test; (**d**) (bottom right) post-test for each conceptual QuILT group.

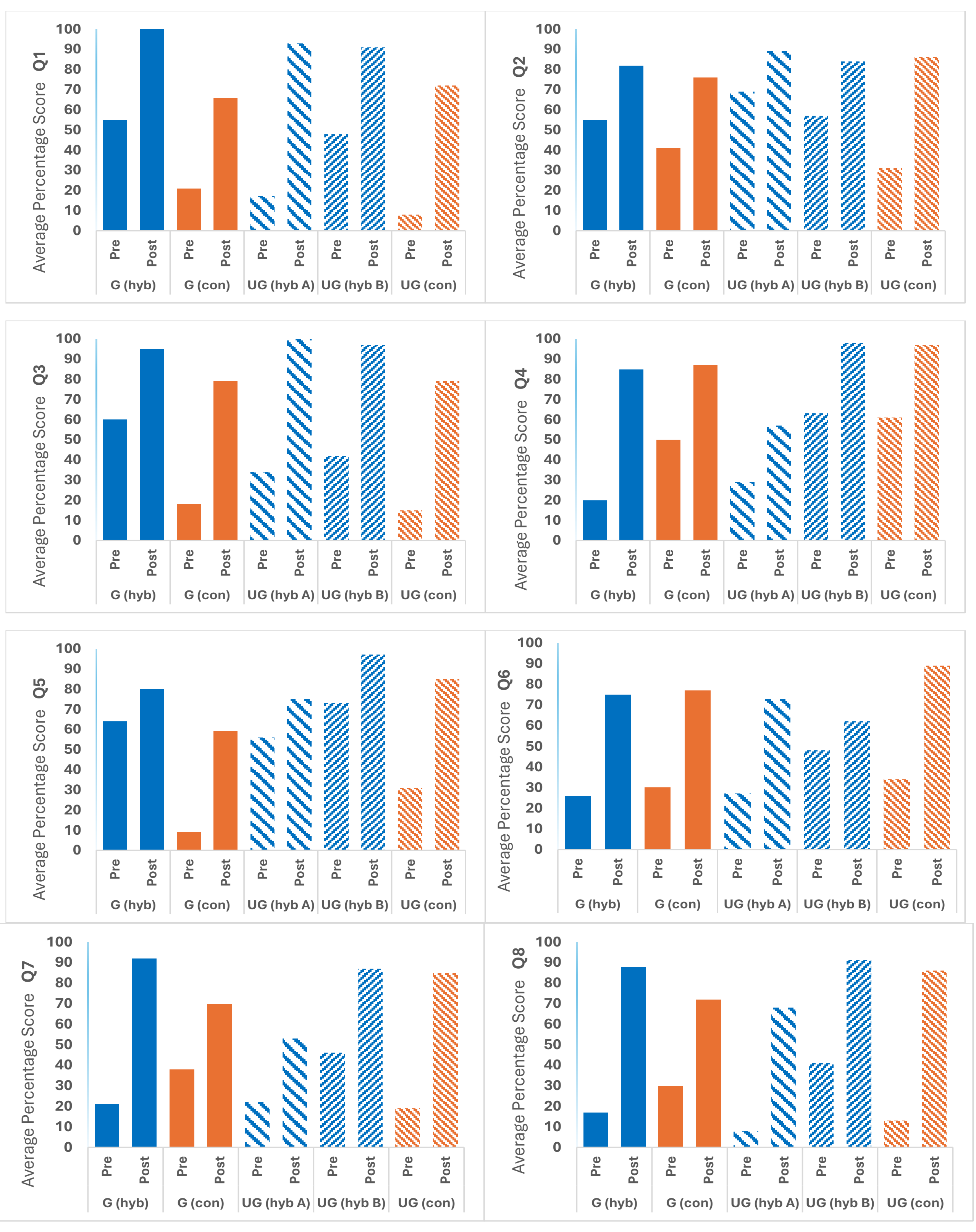

**Figure 6.** Bar graph of average pre-test and post-test scores for Q1–Q8 for each of the five student groups: G (hyb), G (con), UG (hybA), UG (hybB), and UG (con). Abbreviations: graduate (G), undergraduate (UG), hybrid (hyb), conceptual (con).

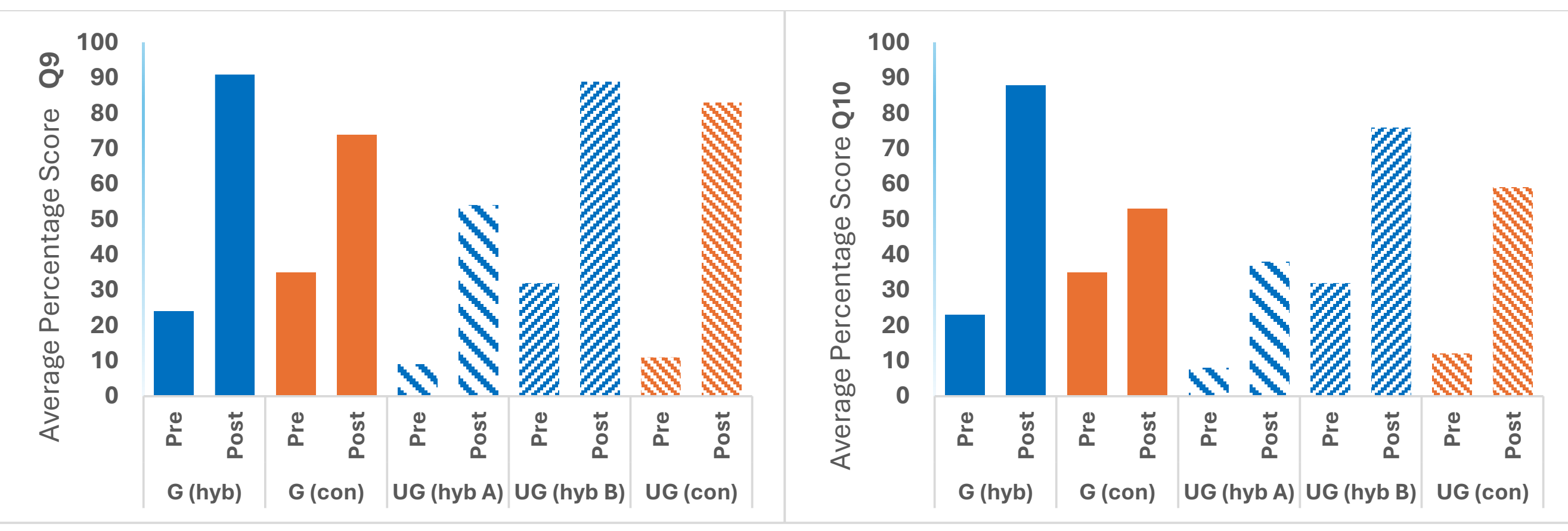

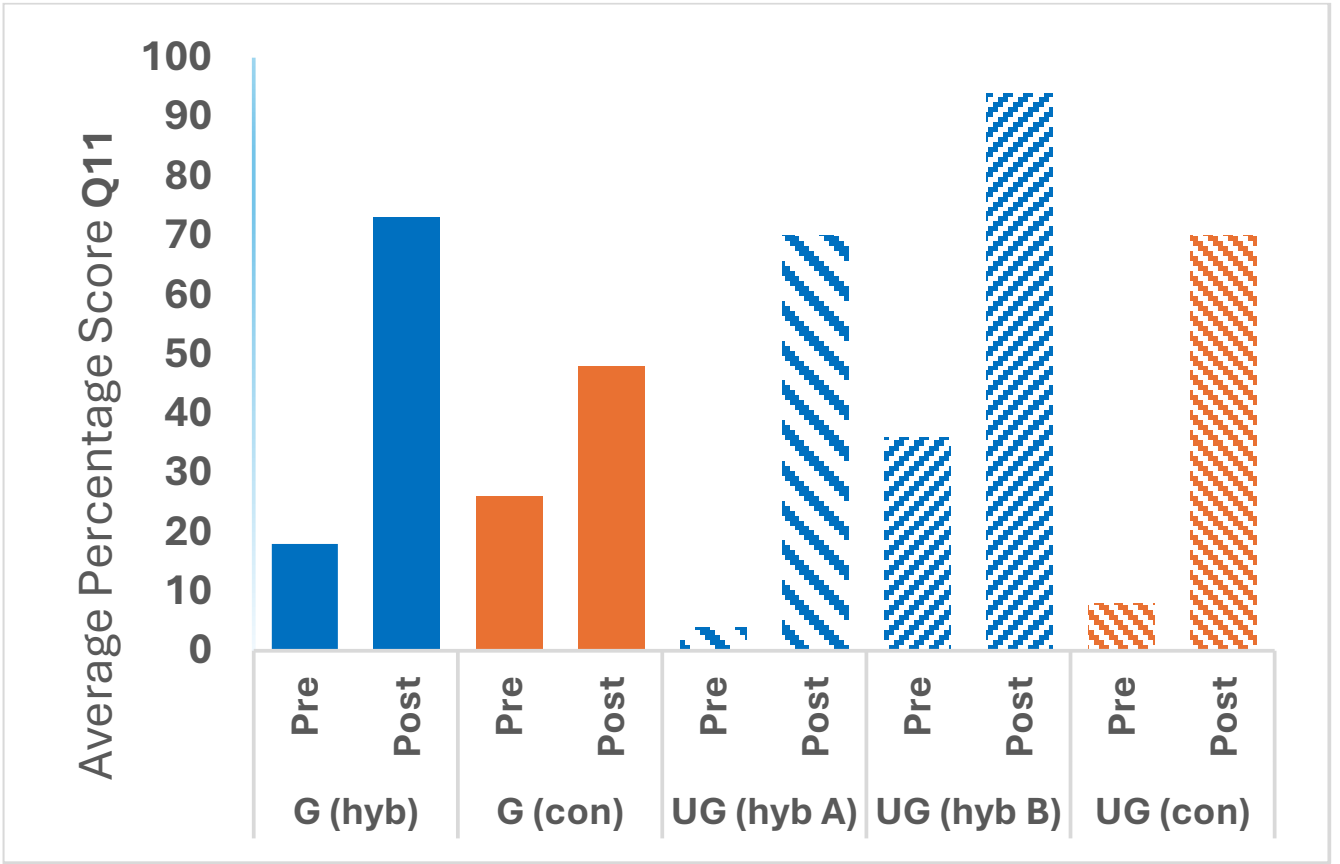

**Figure 7.** Bar graph of average pre-test and post-test scores for Q9–11 for each of the five student groups: G (hyb), G (con), UG (hybA), UG (hybB), and UG (con). Abbreviations: graduate (G), undergraduate (UG), hybrid (hyb), conceptual (con).

### *3.1. RQ1: How Does the Integration of Conceptual and Quantitative Reasoning in a Hybrid QuILT Affect Graduate Students' Conceptual Understanding of Quantum Optics Compared to a Purely Conceptual QuILT?*

A comparison of Figures 6 and 7 (comparison of results in Table 1 with Table A1 in Appendix A from Ref. [96] and Table 2) shows that both groups of graduate students performed somewhat similarly on average on the pre-tests, although there are some variations across questions. For example, the hybrid QuILT group's performance on Q1–Q3 and Q5 suggests their lecture experience made them better prepared for the more basic concepts of the MZI than their conceptual QuILT counterparts. But their performance on some of the other problems shows comparable performance or the conceptual group performing somewhat better on the pre-test for the more complex situations. On the post-test, however, the hybrid QuILT group performed roughly the same as, and in many cases, considerably better than their conceptual QuILT counterpart. One possible explanation is that, for the graduate students, a population that is likely to have higher quantitative facility, at least in the context of physics, the integrated conceptual and quantitative MZI QuILT improves student performance on conceptual questions on some of the concepts compared to the conceptual MZI QuILT.

### *3.2. RQ2: How Does the Integration of Conceptual and Quantitative Reasoning in a Hybrid QuILT Affect Undergraduate Students' Conceptual Understanding of Quantum Optics Compared to a Purely Conceptual QuILT?*

With some variations by problems, Figures 6 and 7 (with results in Table 1, Table A1 from Ref. [96] and Table 2) show that the hybrid QuILT undergraduate group A on average performed somewhat similarly to the conceptual QuILT group in the pre-test. However, for the post-test, undergraduate group A performed better on Q1–3, while performing worse on many of the other problems (detailed comparison of all five groups is in the next section). One possible reason for why undergraduate student group A performed well for Q1–Q3 but not for many of the other problems is attributed to the mathematical facilities of group A (especially in light of the fact that they had poor conceptual understanding of the underlying concepts before engaging with the QuILT as reflected by the low pre-test scores) and the fact that the mathematical rigor in the hybrid QuILT may have caused a cognitive overload for students particularly for the questions involving polarizers that involve product states of path and polarization. In particular, concepts involved in Q1–Q3 require a two-dimensional Hilbert space (only photon path states through the MZI are relevant since polarizers are not present). As such, undergraduates in group A appear to have benefited more from the hybrid QuILT as reflected by their post-test performance. However, for many other problems on the post-test, which involve the four-dimensional Hilbert space (since both photon path states and polarization states must be considered to understand the outcomes of the experiment), their performance is typically worse. On the other hand, undergraduate group B using the hybrid QuILT typically demonstrated stronger pre-test performance compared to the undergraduates who engaged with the conceptual QuILT. One likely reason for this difference is that the instructor of group B may have better prepared and motivated students than the instructor of the conceptual QuILT group. Another possible reason is the differences between students in different years. Furthermore, students in group B typically outperformed all other groups, including those who engaged with the conceptual QuILT, in the conceptual post-test for most of the problems.

### *3.3. RQ3: What Role Does Students' Prior Knowledge Play in Determining Whether They Benefit from the Integrated Conceptual–Quantitative Approach Versus the Conceptual-Only Approach?*

#### 3.3.1. Comparison of Pre-/Post-test Performance of Different Groups That Learned from Hybrid MZI QuILT

Table 1 reveals that both graduate students and the two undergraduate groups performed poorly on the pre-test following traditional instruction. However, after working on the hybrid QuILT, graduate students and undergraduate group B demonstrated strong performance for most post-test questions. A comparison of undergraduate groups A and B shows that group B outperformed group A on most test items in both the pre-test and post-test. In particular, only post-test scores for Q1–Q3, which focus on more basic concepts (where no polarizers are involved and the Hilbert space is two-dimensional), were similar between the two groups.

Also, Table 1 shows that Q4–5 and Q7–11 have considerably higher post-test scores for group B, which tended to also perform better on the pre-test. In the hybrid QuILT, these questions covered concepts and MZI arrangements involving higher mathematical rigor (most involved 4-dimensional space including both path and polarization states as opposed to 2-dimensional space for Q1–Q3). Since these two undergraduate classes were taught at the same university in consecutive years, their populations can be considered comparable in terms of mathematical preparation. One plausible explanation for the post-test score differences between the groups is the disparity in their pre-test scores, which reflects how much they learned about these concepts following lecture-based instruction. The two courses were taught by different instructors, which may have influenced these outcomes. Specifically, Table 1 reveals that undergraduate group A had particularly low pre-test scores on the later questions, likely due to differences in how the instructor for group A covered the material during lectures, the time devoted to the topic before the pre-test, and the extent to which the material was incentivized. It is worth noting that pre-test scores were also very low for some questions even among the graduate students (e.g., see Table 1). However, they still significantly outperformed undergraduate group A on the post-test and performed comparably to undergraduate group B. This is notable because undergraduate group B generally had higher pre-test scores for the later questions than the graduate students.

One possible explanation for the average post-test performance difference between graduate and undergraduate students in group A (e.g., see Table 1), despite both groups having poor pre-test performance, is that graduate students may generally have stronger quantitative skills. This advantage could reduce their cognitive load while learning from the hybrid MZI QuILT, even with their low initial understanding of MZI concepts (as indicated by their pre-test scores). In particular, QM courses at the undergraduate level are required for all students, regardless of whether they plan to pursue graduate studies, leading to a wider range of quantitative proficiency among undergraduates compared to graduate students. It is possible that students with less developed skills to integrate conceptual and quantitative aspects may experience cognitive overload [44] when attempting to integrate both aspects of quantum optics while learning from the hybrid QuILT. This cognitive overload may leave insufficient cognitive resources for sense-making and metacognition. Graduate students, on the other hand, may be better equipped to handle the quantitative demands of the hybrid QuILT, reducing their cognitive load and freeing up cognitive resources for conceptual reasoning and sense-making. Their stronger quantitative facility likely helped mitigate the effects of their initially poor conceptual understanding of the MZI, as evidenced by their pre-test scores. Figure 5a,b supports this hypothesis by showing bar graphs of the pre-test and post-test performance for the three groups that engaged with the hybrid QuILT. While both undergraduate group A and some graduate students performed poorly on the pre-test, many students in undergraduate group A exhibited particularly low pre-test scores after lecture-based instruction. This lack of initial preparation may have contributed to their lower performance in the post-test.

We also note that although our goal is to get all student groups to develop mastery after engaging with the QuILT, since group A had a much lower pre-test score than group B, we can compare hybrid undergraduate groups A and B using normalized gain. Table 1 shows that group A had equal or higher normalized gain in 4 of the 11 questions (Q1–Q3 consisting of 2D Hilbert space and Q6). Since the overall gain of group A is reasonable even though students in group A scored lower on many questions, they did not perform badly considering their initial preparation was weaker via traditional instruction.

#### 3.3.2. Comparison of Performance of All Five Groups Together on Each Question

We now compare the performance of all five groups (graduate and undergraduate groups A and B that engaged with the hybrid QuILT and graduate and undergraduates who engaged with the conceptual QuILT) on each question shown in Figures 6 and 7. Although pre-test performance is included in the figures, here we primarily focus on the comparison of the post-test performance after engaging with the QuILT. However, it is interesting that the pre-test scores varied greatly across different groups. This suggests that even though all instructors covered relevant content via lecture-based instruction before the pre-test, their effectiveness varied.

Regardless of the pre-test performance, Figure 6 shows that all five groups did reasonably well on Q1–Q3 (see Appendix A for the problems), which involve only the two-dimensional vector space consisting of path states, after engaging with either version of the QuILT. For example, Q3 focuses on the role of beamsplitter 2 (BS2) and how it affects interference of single photons. This question investigates whether students understand how removing or inserting beamsplitter BS2 will change the probability of the single photons arriving at each detector D1 or D2. The integrated conceptual and quantitative inquiry-based sequences involving a two-dimensional Hilbert space were designed to provide scaffolding support to have students contemplate the role of BS2 and whether interference is observed for single photons without BS2. The post-test performance on Q3 of all student groups after engaging with the hybrid QuILT is close to perfect, whereas both graduate and undergraduate students averaged 79% after engaging with the conceptual QuILT (see Figure 6).

We note that Q4 (see Appendix A) evaluates student understanding of how adding a detector in the U path affects the interference at D1 and D2. Students need to understand that adding an additional detector would collapse the photon's state into either the U or L path state, rather than the photon remaining in a superposition of both U and L path states. They must also understand how this collapse causes the detectors D1 or D2 after BS2 to click with equal probability and disrupts the interference at D1 and D2. As noted earlier, this type of situation is covered in the conceptual QuILT but is not covered in the hybrid QuILT, although students had learned about the role of detectors D1 and D2 after BS2. Figure 6 shows that on the post-test, graduate students and undergraduates in group B who engaged with the hybrid QuILT performed comparably to the graduate and undergraduate students in the conceptual QuILT group. This is encouraging since it suggests that students in these hybrid groups were able to transfer their learning about the role of detectors from those after BS2 (photodetectors D1 and D2) to a detector in the U path of the MZI. Only undergraduate student group A engaging with the hybrid QuILT performed poorly on Q4 on the post-test.

For the questions involving polarizers, students were told to assume that the source emits +45° polarized single photons. Q5 in Appendix A asks students to evaluate the validity of a statement involving a source emitting +45° polarized single photons and whether placing additional polarizers in the paths of the MZI will affect whether interference is displayed at the detectors. We wanted students to describe that polarizers can affect interference and that photon path states which are orthogonally polarized cannot show interference. Figure 6 shows that on the post-test, the undergraduate student group B using the hybrid QuILT performed the best and the only group that performed below 60% on Q5 was the graduate student group using the conceptual QuILT (which we hypothesize could be due to the fact that all graduate students may not have engaged meaningfully in answering this question which did not have a concrete scenario since the Teaching of Physics course is a pass/fail course).

Q6 in Appendix A asks students about the fraction of N photons reaching the detectors D1 and D2 and showing interference when a polarizer with a vertical polarization axis is inserted into the original set up in Q1. This problem is challenging since students must recognize that N/4 photons are absorbed by the vertical polarizer. Out of the rest of the single photons, N/2 + N/8 would arrive at the detector D1 and N/8 would arrive at the detector D2, and it is the N/2 photons with the vertical polarization component that show interference (Supplementary Materials describe this situation and the fact that which-path information (WPI) is not known about these photons). Regardless of the pre-test performance, Figure 6 shows that the undergraduate conceptual group performed better on the post-test than all hybrid groups, suggesting this was a challenging situation for all hybrid groups.

Q7 in Appendix A asks students about the fraction of photons reaching the detectors D1 and D2 and showing interference when two polarizers with horizontal and vertical polarization axes are inserted into the lower and upper paths, respectively, of the original set up in Q1. In this problem, students must recognize that N/4 photons are absorbed by each of the horizontal and vertical polarizers. Out of the remaining single photons, N/4 would arrive at detector D1 and N/4 would arrive at detector D2 and none of the photons show interference (Supplementary Materials describe this situation and the fact that which-path information or WPI is known about all photons), so putting a phase shifter in one of the paths would not affect outcomes. Figure 6 shows that the graduate students and undergraduate group B using the hybrid QuILT and the undergraduates using the conceptual QuILT performed very well after engaging with the QuILT (85% or above for all these groups) regardless of their pre-test performance. However, undergraduate group A using the hybrid QuILT performed relatively poorly on the post-test with an average score of 53%.

Q8 in Appendix A is the quantum eraser set up and asks students about the fraction of photons reaching the detectors D1 and D2 and showing interference when two polarizers with horizontal and vertical polarization axes are inserted into the lower and upper paths, respectively, of the original set up in Q1 and an additional polarizer with +45° polarization axis is inserted between beamsplitter 2 and detector D1. In this problem, students must recognize that N/4 photons are absorbed by each of the horizontal and vertical polarizers. Out of the rest of the single photons, N/4 would arrive at the detector D1 and would show interference (the additional polarizer with +45° before the detector D1 is a quantum eraser and erases the which-path information so that the photon path states from the upper and lower paths can interfere) and N/4 that arrive at the detector D2 would not show interference (Supplementary Materials describe this situation and the fact that which-path information is not known for the photons arriving at D1 but it is known for those arriving at D2). In this situation, putting a phase shifter in one of the paths would affect the photons that show interference. Figure 6 shows that the graduate students and undergraduate group B using the hybrid QuILT and the undergraduates using the conceptual QuILT performed very well after engaging with the QuILT (85% or above for all these groups) regardless of their pre-test performance. However, the undergraduate group A using the hybrid QuILT and graduate students using the conceptual QuILT performed comparably on the post-test with an average approximate score of 70%.

Q9 in Appendix A asks students about the fraction of photons reaching the detectors D1 and D2 and showing interference when two polarizers with horizontal and vertical polarization axes are inserted into the lower and upper paths, respectively, of the original set up in Q1 and a third polarizer with horizontal polarization axis is inserted between beamsplitter 2 and detector D1. In this problem, students must recognize that N/4 photons are absorbed by each of the horizontal and vertical polarizers in the upper and lower paths of the MZI. Out of the remaining single photons, N/8 would be absorbed by the third polarizer between the beamsplitter 2 and detector D1 and the other N/8 would arrive at the detector D1 but they would not show interference (the additional polarizer with horizonal polarization before the detector D1 cannot erase the which-path information so that the photon path states from the upper and lower paths cannot interfere) and N/4 that arrive at the detector D2 would also not show interference (in other words, which-path information is known for all photons arriving at detectors D1 or D2). In this situation, putting a phase shifter in one of the paths would not affect the photons since none of them would show interference in this case. Figure 7 shows that the graduate students and undergraduate group B using the hybrid QuILT obtained an average score of approximately 90% on post-test but undergraduate group A performed the worst at 54%. The graduate and undergraduate conceptual groups had average post-test scores of 74% and 83%, respectively. This problem can be considered a near transfer problem since students using both hybrid and conceptual versions of the QuILT had worked through a situation in which the polarizer right before the detector D1 was a vertical polarizer (instead of a horizontal polarizer).

Q10 in Appendix A is a far transfer problem (because there was no such situation in the QuILT with two polarizers between the beamsplitter 2 and detector D1) and asks students about the fraction of photons reaching the detectors D1 and D2 and showing interference when two polarizers with horizontal and vertical polarization axes are inserted into the lower and upper paths of the original set up in Q1 and the third and fourth polarizers (with horizontal polarization and +45° polarization axes) are inserted one after another between the beamsplitter 2 and detector D1. In this problem, students must recognize that N/4 photons are absorbed by each of the horizontal and vertical polarizers in the upper and lower paths of the MZI. Out of the remaining single photons, N/8 would be absorbed by the third polarizer with the horizontal polarization between the beamsplitter 2, and detector D1 and another N/16 will be absorbed by the polarizer with +45° polarization axis. The other N/16 would arrive at detector D1, but they would not show interference (the two additional polarizers between beamsplitter 2 and the detector D1 cannot erase the which-path information so that the photon path states from the upper and lower paths cannot interfere). Also, N/4 photons that arrive at detector D2 would also not show interference (in other words, in this situation, which-path information is known for all photons arriving at detectors D1 or D2). In this situation, putting a phase shifter in one of the paths would not affect the photons since none of them show interference in this case. Figure 7 shows that this far transfer problem was challenging and only the graduate students and undergraduate group B using the hybrid QuILT obtained an average score greater than 75% on the post-test while the other three groups performed below 60%. The major difference between the graduate students and undergraduate group B using the hybrid QuILT compared to the graduate and undergraduate conceptual groups in this far transfer problem may be due to the benefits of the integrated conceptual and quantitative QuILT for deeper learning.

Q11 asks students to describe an experiment using an MZI in which one could distinguish between a source emitting unpolarized photons and a source emitting +45° polarized photons. There are many possible correct responses for this, e.g., students could have a quantum eraser set up that only erases which-path information for polarized photons so that only that case interference would be observed at detectors D1 and D2, or one could have a set up in which a polarizer with −45° polarization axis is placed in the lower path of the MZI. In this case, if the source is unpolarized, it would have some photons showing interference and others not showing interference. However, if the source emits +45° polarized single photons, the −45° polarizer would block them all so we have which-path information for all photons arriving at detectors D1 and D2, and no interference is observed. Figure 7 shows that this problem was challenging for some groups even for the post-test, and while undergraduate group B using hybrid QuILT performed better than 90%, the graduate group and undergraduate group A using the hybrid QuILT and the undergraduate conceptual group performed close to 70%.

The question-by-question analysis reveals a clear pattern on how different groups benefited from the hybrid versus conceptual QuILT versions. For questions involving only two-dimensional Hilbert space (Q1–Q3, which deal with basic MZI concepts without polarizers), all student groups performed well on the post-test regardless of which QuILT version they used, with scores typically above 80%. However, a stark divide emerged for questions requiring an understanding of four-dimensional Hilbert space involving polarizers and product states of path and polarization. On many of these more complex questions, graduate students and undergraduate group B using the hybrid QuILT typically outperformed or matched the conceptual QuILT groups, while undergraduate group A using the hybrid QuILT generally performed worse.

This pattern suggests that the benefits of integrating quantitative and conceptual reasoning depend heavily on students' prior preparation and mathematical facility. We hypothesize that graduate students, with their stronger quantitative background, were able to leverage the mathematical framework in the hybrid QuILT to develop deeper conceptual understanding, even for complex scenarios involving four-dimensional Hilbert space. Similarly, undergraduate group B, who demonstrated better conceptual understanding in the pre-test, could effectively use quantitative tools to enhance their learning without experiencing cognitive overload. However, undergraduate group A, which showed poor pre-test performance, appeared to experience cognitive overload when dealing with mathematical frameworks in four-dimensional space, leading to worse conceptual understanding compared to students who used the purely conceptual QuILT. The only notable exception to this pattern was Q6 (involving a single polarizer), where the undergraduate conceptual QuILT group outperformed all hybrid QuILT groups.

## 4. Broader Discussion

### *4.1. Interpretation of Results Through ICQUIP Framework*

The findings strongly support the ICQUIP framework's central premise that successful integration of conceptual and quantitative aspects requires careful consideration of students' prior knowledge and cognitive load management. The differential effectiveness of the hybrid QuILT across student groups in achieving mastery as measured by their post-test performance illustrates several key principles. For example, consistent with our hypothesis H1, our findings suggest that graduate students and well-prepared undergraduates (group B) could manage cognitive load and possess sufficient cognitive resources to process mathematical formalism while maintaining focus on conceptual understanding. Their stronger foundation prevented cognitive overload, allowing the quantitative tools to serve as scaffolds for deeper sense-making. Regarding our hypothesis H2 pertaining to prior knowledge requirements, our findings suggest that effective use of the hybrid QuILT appears to require strong mathematical facility with linear algebra including handling of 4x4 matrices, basic understanding of quantum superposition and measurement, familiarity with polarization concepts in the context of MZI and ability to translate between mathematical representations and physical meaning. Students lacking these prerequisites (e.g., in undergraduate group A) may have experienced cognitive overload. Consistent with H3, this was especially true for problems involving 4D Hilbert space. These findings suggest that assessment of prior knowledge should precede implementation of integrated conceptual and quantitative approaches.

Our findings align with international studies on the integration of mathematical formalism and abstract tools in physics learning. For example, a German study on quantum physics education[75], which was conducted in the context of quantum optics, was summarized as follows by the authors, "According to common sense, a teaching concept mainly based on a research paper is likely to be too advanced for high school pupils. The present investigation shows that this is no problem in the present case. The pupils can cope with central concepts including coincidence counts and anticorrelation". Similar to our findings for some student groups, in their study, students were able to handle the complexity of challenging concepts such as coincidence counts and anticorrelation potentially because the scaffolding provided was commensurate with their prior knowledge. The importance of metacognitive support in our work echoes the work from Italy on teaching quantum mechanics with Dirac notation[58].

For students with insufficient prior knowledge, several interventions could improve readiness for the hybrid QuILT. For example, pre-instruction modules that provide students with opportunities to review relevant mathematics could be helpful. Also, diagnostic assessments to identify specific gaps and differentiated pathways through the QuILT based on prior knowledge could be valuable. Additional scaffolding for transitions between 2D and 4D representations may be particularly useful for students who struggled with that transition.

### 4.2. Limitations

Several limitations should be considered when interpreting these results. The small sample sizes limit statistical power and generalizability. Also, all data are from a single university, which may not represent broader populations of physics students. Also, different instructors with their own traditional lecture approaches may have influenced students' initial preparation levels. Also, using identical pre-/post-tests may introduce testing effects even though the questions on these assessments were open-ended, so the testing effect is unlikely to be a major influence. Also, implementation variations, e.g., pass/fail grading for some groups may have affected engagement with the QuILT and pre-/post-test.

### 4.3. Future Directions

Future research should investigate optimal prior knowledge thresholds to benefit from integrated conceptual and quantitative approaches, develop adaptive versions that adjust mathematical complexity based on students' readiness, examine long-term retention and transfer of integrated versus conceptual-only learning, and explore the framework's applicability to other physics domains beyond quantum optics. Similar investigations at other institutions of different types within a given country as well as across different countries around the world would be very valuable.

## 5. Conclusions

This study used the ICQUIP framework to investigate how integrating conceptual and quantitative reasoning affects students' understanding of quantum optics. The results demonstrate that such integration can enhance learning when properly calibrated to students' prior knowledge and mathematical facility. Graduate students consistently benefited from the hybrid approach, while undergraduate students' benefits, regarding achieving a level of mastery, depended critically on their initial preparation level. These findings underscore that "one size does not fit all" in physics education. The ICQUIP framework provides guidance for when and how to integrate mathematical formalism including ensuring students have sufficient prior knowledge, providing appropriate scaffolding, and creating opportunities for metacognitive reflection on the connections between mathematics and physics.

The study contributes to our understanding of how to design effective research-based learning tools that leverage the power of mathematical representation while avoiding cognitive overload. As physics education continues to evolve, frameworks like ICQUIP will be essential for creating learning environments that help all students develop integrated conceptual–quantitative expertise that is characteristic of physics mastery.

**Supplementary Materials:** The following supporting information can be downloaded at https://www.mdpi.com/article/doi/s1, and provide a summary of development and validation of the two versions of the MZI QuILT, a brief background and how the QuILT addresses common difficulties.

**Author Contributions:** All authors contributed equally to all aspects of this research. All authors have read and agreed to the published version of the manuscript.

**Funding:** This research is supported by the US National Science Foundation Award PHY-2309260.

**Institutional Review Board Statement:** This research was carried out in accordance with the principles outlined in the University of Pittsburgh Institutional Review Board (IRB) ethical policy, the Declaration of Helsinki, and local statutory requirements.

**Informed Consent Statement:** Informed consent was obtained from students who were interviewed individually.

**Data Availability Statement:** The datasets used and analyzed during the current study are not available due to confidentiality agreement with interviewees.

**Conflicts of Interest:** The authors declare no conflict of interest.

## Appendix A

### Table from Ref. [96]

Table A1. Percentages of undergraduates and graduate students who correctly answered questions on the MZI pre-test/post-test before and after using the conceptual-only QuILT after averaging over the sub-parts of each question and normalized gains. Data were primarily taken from Ref. [96], with sub-questions averaged for comparison (we note that a few question numberings differed from the prior study but have been matched appropriately). Unless otherwise specified, the number of graduate students was 45 (matched), the numbers of undergraduates were 44 (pre) and 38 (post), with data collected over a period of two years in both cases. (We note that some problem numbers differ from Ref.[96], i.e., problems (3), (4), (5), (6), and (7) in Ref. [96]are problems (3a), (3b), (4), (7), and (8), respectively, in Appendix A).

| | Graduate Students | | | Undergraduates | | |
|---|---|---|---|---|---|---|
| Q | Pre(%) | Post(%) | <g> | Pre(%) | Post(%) | <g> |
| 1 | 21 | 66 | 0.57 | 8 | 72 | 0.70 |
| 2 | 41 | 76 | 0.59 | 31 | 86 | 0.80 |
| 3 | 18 | 79 | 0.74 | 15 | 79 | 0.75 |
| 4 | 50 | 87 | 0.74 | 61 | 97 | 0.92 |
| 7 | 38 | 70 | 0.52 | 19 | 85 | 0.81 |
| 8 | 30 | 72 | 0.60 | 13 | 86 | 0.84 |

### Mach–Zehnder Interferometer (MZI) Pre-/Post-test

(Note: More space was provided for students to answer questions.)

The set up for the ideal Mach–Zehnder Interferometer (MZI) shown below in Figure A1 is as follows:

- The photons originate from a monochromatic coherent point source. (Note: Experimentally, a source can only emit nearly monochromatic photons such that there is a very small range of wavelengths coming from the source. Here, we assume that the photons have negligible “spread” in energy.)
- Assume that the photons propagating through both the U and L paths travel the same distance in vacuum to reach each detector.
- All angles of incidence are 45° with respect to the normal to the surface.
- For simplicity, we will assume that a photon can only reflect from one of the two surfaces of the identical half-silvered mirrors (beamsplitters) BS1 and BS2 because of an anti-reflection coating on one of the surfaces.
- Assume that beamsplitters BS1 and BS2 are infinitesimally thin so that there is no phase shift when a photon propagates through them.
- The phase shifter is ideal and non-reflective.

- Ignore the effect of polarization of the photons due to reflection by the beamsplitters or mirrors.
- The photodetectors D1 and D2 are point detectors located symmetrically with respect to the other components of the MZI as shown.
- All photodetectors are ideal and 100% efficient.
- Polarizers do not introduce phase shifts.
- All measurements are ideal projective measurements.

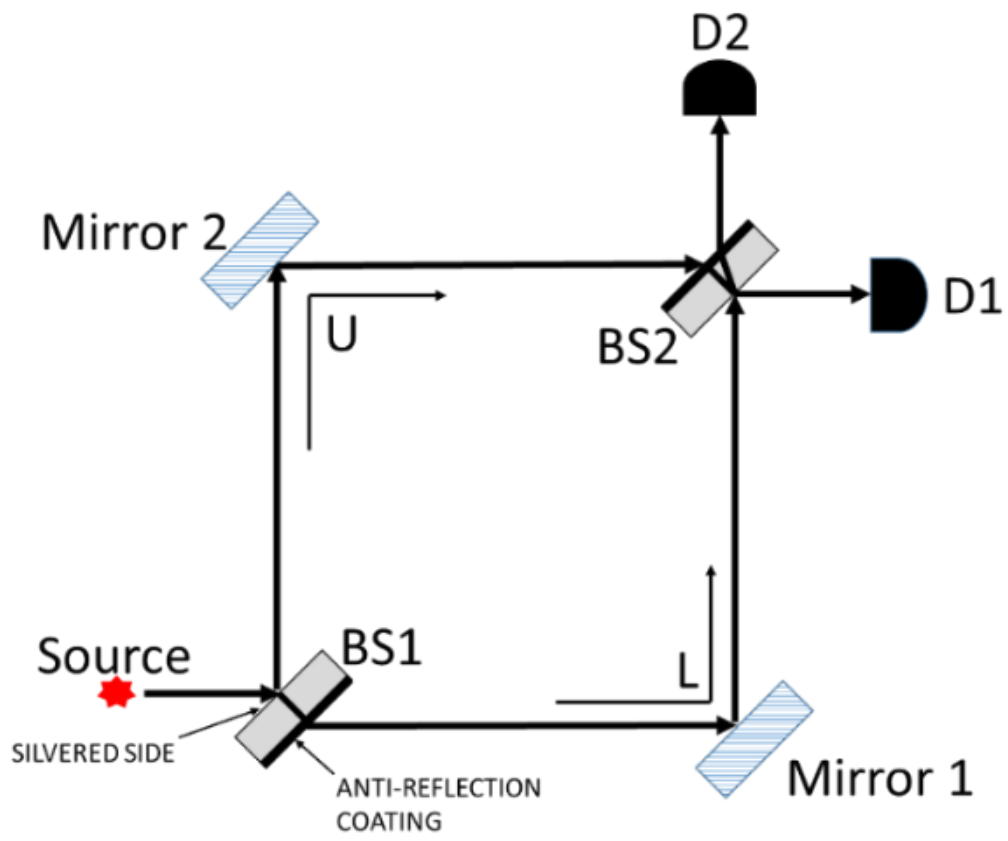

**Figure A1.** Basic setup of the MZI interferometer

For all of the following questions, assume that

- The single photons are emitted from the source in a highly collimated stream, i.e., the width of the transverse Gaussian profile of each photon is negligible.
- A very large number (N) of single photons are emitted from the source one at a time and passes through beamsplitter BS1.

1. Consider the following statement about single photons emitted from the source in Figure A1:
   - If the source emits N photons one at a time, the number of photons reaching detectors D1 and D2 will be N/2 each.

   Explain why you agree or disagree with this statement.
2. Consider the following conversation between Student 1 and Student 2:
   - Student 1: The beamsplitter BS1 causes the photon to split into two parts, and the energy of the incoming photon is also split in half. Each photon with half of the energy of the incoming photon travels along the U and L paths of the MZI and produces interference at detectors D1 and D2.
   - Student 2: If we send one photon at a time through the MZI, there is no way to observe interference in the detectors D1 and D2. Interference is due to the superposition of waves from the U and L paths. A single photon must choose either the U or the L path.

   Do you agree with Student 1, Student 2, both, or neither? Explain your reasoning.

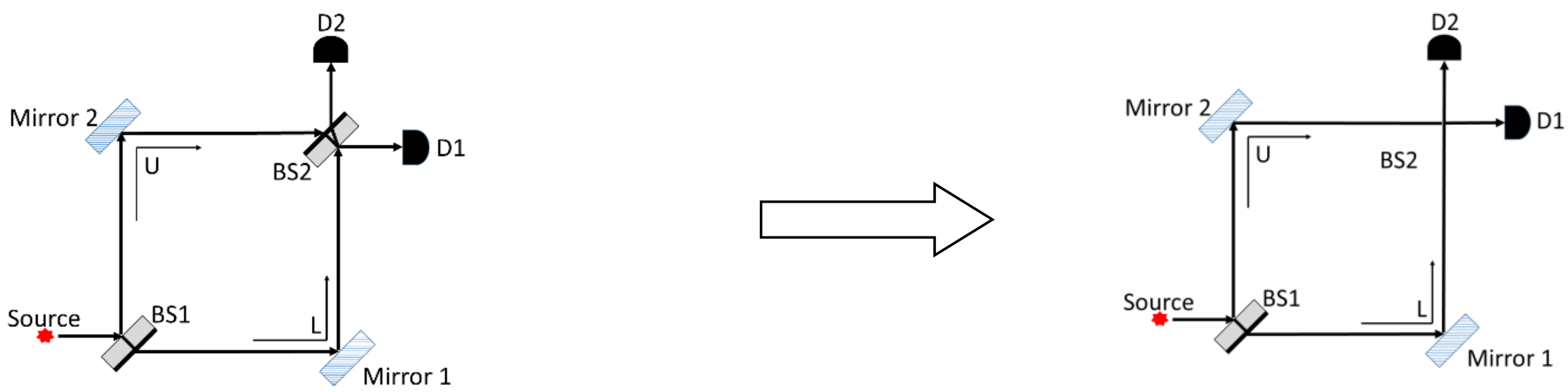

**Figure A2.** Changes in the setup described in Question 3a

3a. Suppose we remove BS2 from the MZI set up as shown in Figure A2 above. How does the probability that detector D1 or D2 will register a photon in this case differ from the case when BS2 is present as in Figure A1? Explain your reasoning.

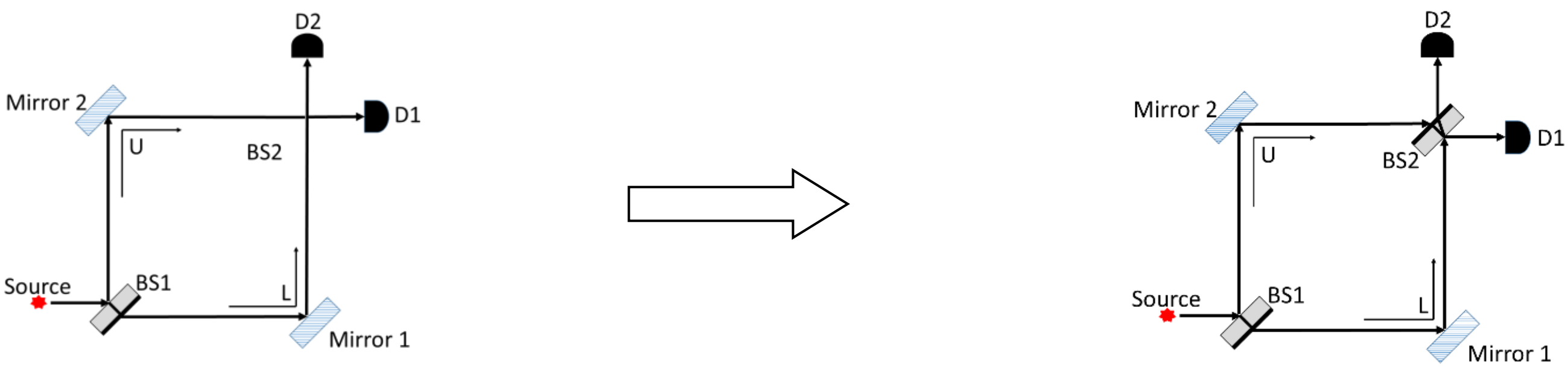

**Figure A3.** Changes in the setup described in Question 3b

3b. Suppose we have an MZI set up initially without BS2. If we suddenly insert BS2 after the photon enters BS1 but before it reaches the point where BS2 is inserted (see Figure A3), with what probabilities do detectors D1 and D2 register the photon? Explain your reasoning. Assume that the situation after BS2 is inserted is identical to Figure A1.

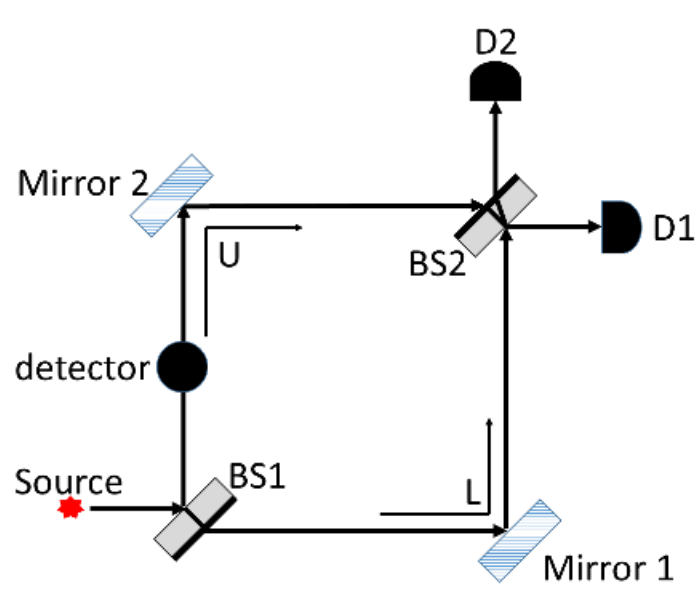

**Figure A4.** Setup for Question 4

4. Suppose we modify the setup shown in Figure A1 and insert a photodetector into the upper path between BS1 and mirror 2 as shown in Figure A4.
   (a) What is the fraction of single photons emitted by the source that reach each detector D1 and D2? Explain your reasoning.
   (b) If you place a phase shifter in the L path and change its thickness gradually to change the path length difference between the U and L paths, how would the phase shifter affect the fraction of photons arriving at detectors D1 and D2? Explain your reasoning.

(c) If there is interference displayed in part 4(b) by any photons at detector D1, write down the percentage of the photons emitted by the source that display interference. You must explain your reasoning.

For all of the following questions, assume that the single-photon source emits photons that are polarized at +45°.

5. Consider the following statement about a source emitting +45° polarized single photons:

   - If we place additional polarizers in the paths of the MZI, the polarizers will absorb some photons and they will not arrive at the detectors. However, the polarizers will not affect whether interference is displayed at the detectors.

   Explain why you agree or disagree with the statement.

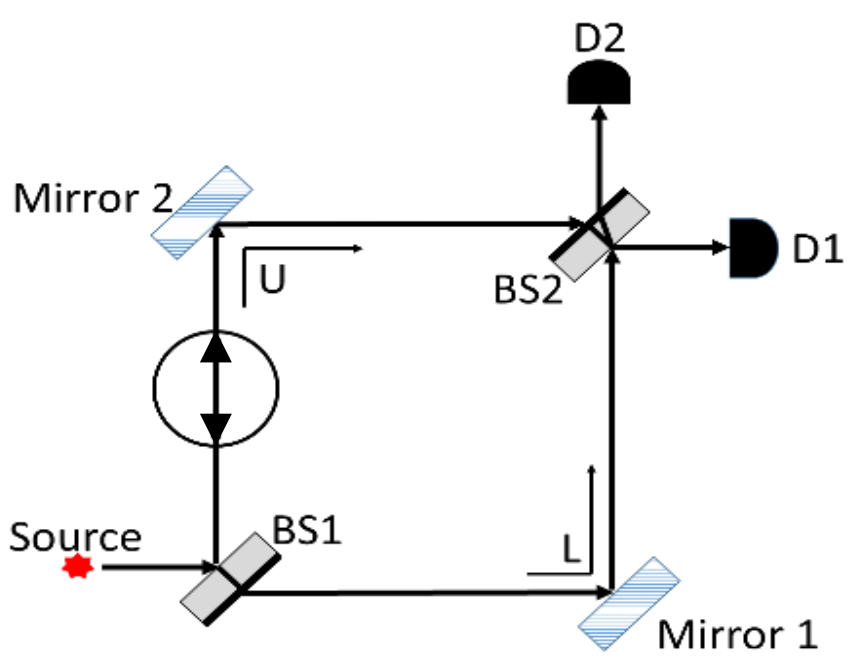

**Figure A5.** Setup for Question 6

6. You modify the setup shown in Figure A1 by inserting a polarizer with a vertical polarization axis as shown in Figure A5.

   (a) What is the fraction of single photons emitted by the source that reach each detector D1 and D2? Explain your reasoning.

   (b) If you place a phase shifter in the U path and change its thickness gradually to change the phase difference between the U and L paths, how would the phase shifter affect the fraction of photons arriving at detectors D1 and D2? Explain your reasoning.

   (c) If there is interference displayed by any photons in part 6(b) at detector D1, write down the percentage of the photons emitted by the source that displays interference. You must explain your reasoning.

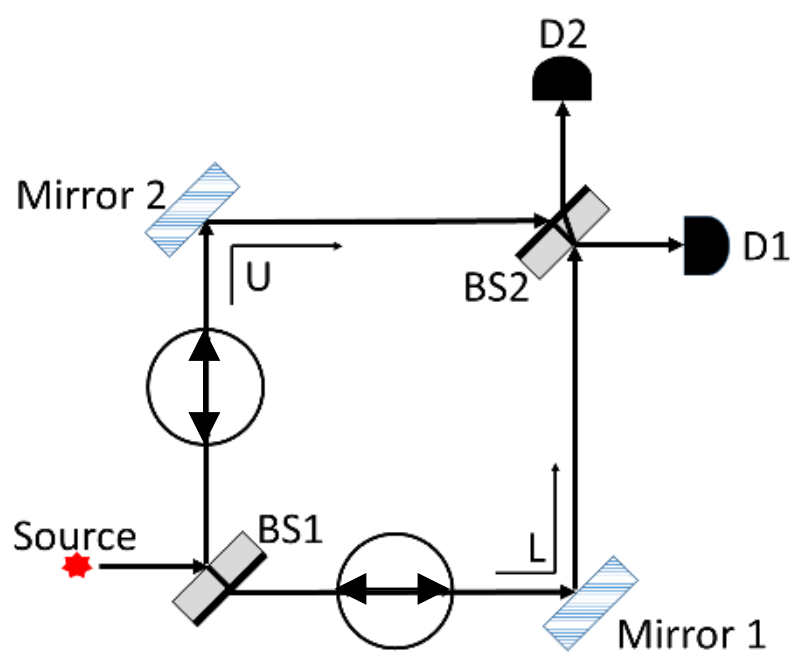

**Figure A6.** Setup for Question 7

7. You modify the setup shown in Figure A1 and insert polarizer 1 with a vertical polarization axis (between BS1 and mirror 2) and polarizer 2 with a horizontal polarization axis (between BS1 and mirror 1) in the U and L paths as shown in Figure A6.

(a) What is the fraction of single photons emitted by the source that reaches each detector D1 and D2? Explain your reasoning.

(b) If you place a phase shifter in the U path and change its thickness gradually to change the phase difference between the U and L paths, how would the phase shifter affect the fraction of photons arriving at detectors D1 and D2? Explain your reasoning.

(c) If there is interference displayed by any photons in part 7(b) at detector D1, write down the percentage of the photons emitted by the source that displays interference. You must explain your reasoning.

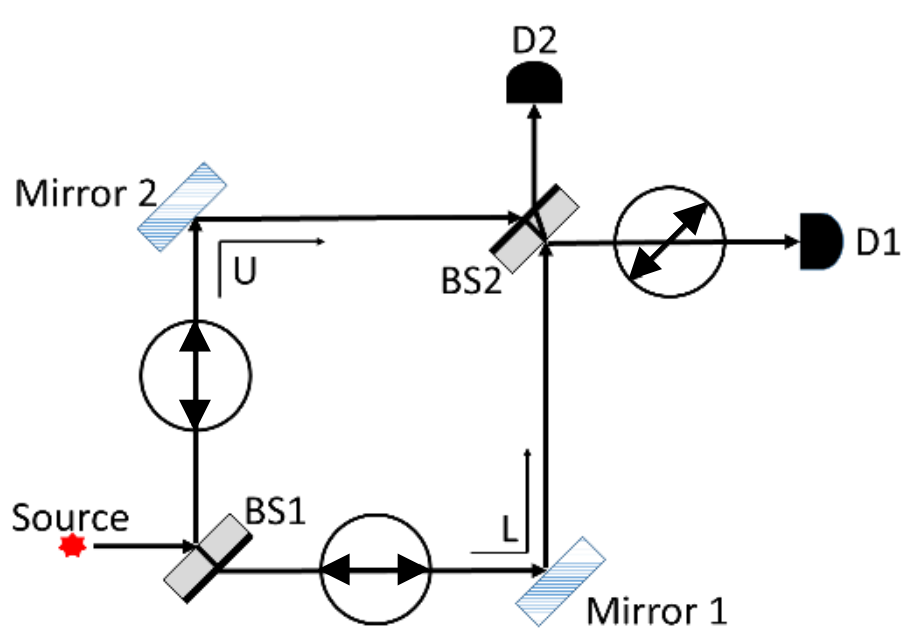

**Figure A7.** Setup for Question 8

8. You start with the setup shown in Figure A6 with polarizer 1 with a vertical polarization axis and polarizer 2 with a horizontal polarization axis inserted in the U and L paths, respectively. You modify the set up and insert polarizer 3 with a +45° polarization axis between BS2 and detector D1 (see Figure A7).

   (a) What is the fraction of single photons emitted by the source that reaches each detector D1 and D2? Explain your reasoning.

   (b) If you place a phase shifter in the U path and change its thickness gradually to change the phase difference between the U and L paths, how would the phase shifter affect the fraction of photons arriving at detectors D1 and D2? Explain your reasoning.

   (c) If there is interference displayed by any photons in part 8(b) at detector D1, write down the percentage of the photons emitted by the source that displays interference. You must explain your reasoning.

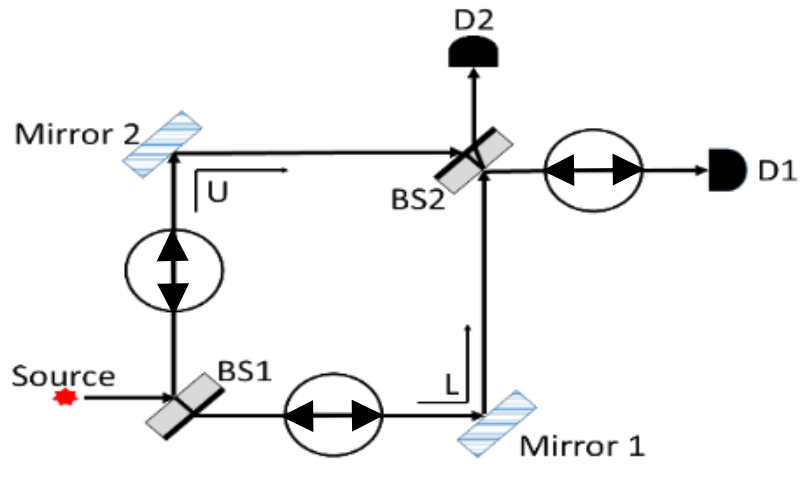

**Figure A8.** Setup for Question 9

9. You start with the setup shown in Figure A6 with polarizer 1 with a vertical polarization axis and polarizer 2 with a horizontal polarization axis inserted in the U and L paths, respectively. You modify the set up and insert polarizer 3 with a horizontal polarization axis between BS2 and the detector D1 (see Figure A8).

   (a) What is the fraction of single photons emitted by the source that reaches each detector D1 and D2? Explain your reasoning.

(b) If you place a phase shifter in the U path and change its thickness gradually to change the phase difference between the U and L paths, how would the phase shifter affect the fraction of photons arriving at detectors D1 and D2? Explain your reasoning.

(c) If there is interference displayed by any photons in part 9(b) at detector D1, write down the percentage of the photons emitted by the source that displays interference. You must explain your reasoning.

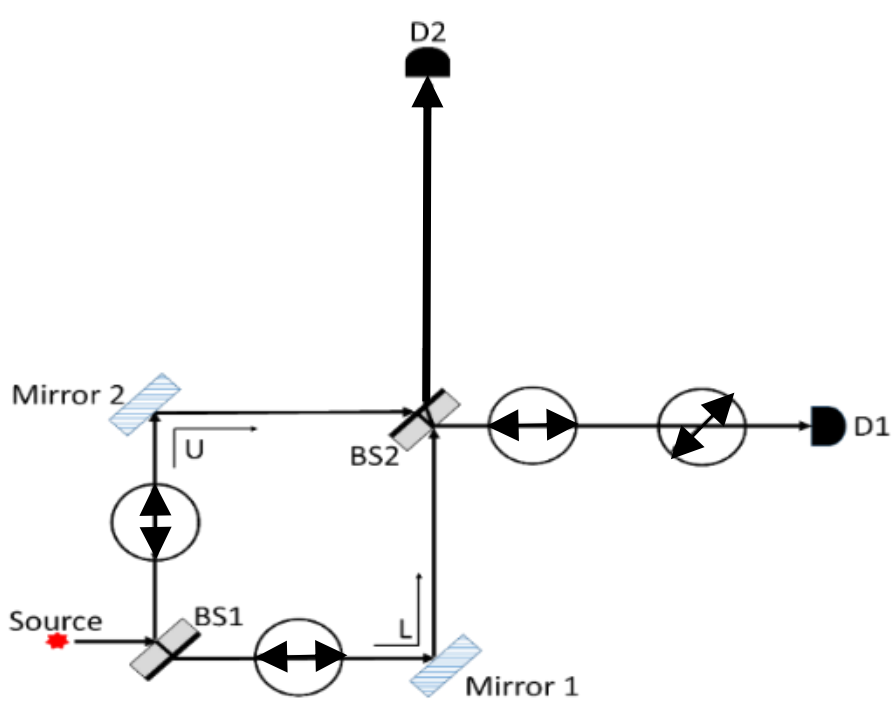

**Figure A9.** Setup for Question 10

10. You set up an MZI as shown in Figure A9, inserting polarizer 1 with a vertical polarization axis and polarizer 2 with a horizontal polarization axis in the U and L paths, respectively. You also insert polarizer 3 with a horizontal polarization axis and polarizer 4 with a 45° polarization axis between BS2 and detector D1 (see Figure A9).

    (a) What is the fraction of single photons emitted by the source that reaches each detector D1 and D2? Explain your reasoning.

    (b) If you place a phase shifter in the U path and change its thickness gradually to change the phase difference between the U and L paths, how would the phase shifter affect the fraction of photons arriving at detectors D1 and D2? Explain your reasoning.

    (c) If there is interference displayed by any photons in part 10(b) at detector D1, write down the percentage of the photons emitted by the source that displays interference. You must explain your reasoning.

11. Describe an experiment using a Mach–Zehnder Interferometer in which you could distinguish between a source emitting unpolarized photons and a source emitting +45° polarized photons.

## References

[1] Sternberg R J 1998 Metacognition, abilities, and developing expertise: What makes an expert student? *Instructional Science* **26** (1) 127-140

[2] diSessa A A (1988). Knowledge in pieces. Constructivism in the computer age. G. Forman and P. Pufall. Hillsdale, NH, Lawrence Erlbaum**:** 49-70.

[3] diSessa A A 1993 Toward an epistemology of physics. *Cognition and Instruction* **10** (2-3) 105-225

[4] Berliner D (1994). Expertise: The wonder of exemplary performances. Creating powerful thinking in teachers and students: Diverse perspectives. Ft. Worth, TX, Holt, Rinehart and Winston**:** 141-186.

[5] Hammer D 1994 Epistemological beliefs in introductory physics. *Cognition and Instruction* **12** (2) 151-183

[6] Elby A 2001 Helping physics students learn how to learn. *Am. J. Phys.* **69** (S1) S54--S64

[7] DiSessa A A (2018). A friendly introduction to “knowledge in pieces”: Modeling types of knowledge and their roles in learning. Invited lectures from the 13th International Congress on Mathematical Education, Springer.

[8] Schraw G 1998 Promoting general metacognitive awareness. *Instructional Science* **26** (1) 113-125

[9] Schoenfeld A H 2016 Learning to think mathematically: Problem solving, metacognition, and sense making in mathematics. *The Journal of Education* **196** (2) 1-38

[10] Newell A and Simon H A (1972). Human problem solving. Englewood Cliffs, NJ, Prentice-Hall.

[11] Kotovsky K, et al. 1985 Why are some problems hard? Evidence from tower of hanoi. *Cognitive Psychology* **17** 248-294

[12] Ericsson K A and Smith J (1991). Toward a general theory of expertise: Prospects and limits. New York, NY, US, Cambridge University Press.

[13] Chi M T H, et al. 1981 Categorization and representation of physics problems by experts and novices. *Cognitive Science* **5** (2) 121--152

[14] Bing T J and Redish E F (2007). The cognitive blending of mathematics and physics knowledge. AIP Conference Proceedings, American Institute of Physics.

[15] Tuminaro J and Redish E F 2007 Elements of a cognitive model of physics problem solving: Epistemic games. *Physical Review Special Topics-Physics Education Research* **3** (2) 020101

[16] Bing T J and Redish E F 2009 Analyzing problem solving using math in physics: Epistemological framing via warrants. *Physical Review Special Topics - Physics Education Research* **5** (2) 020108

[17] Bing T J and Redish E F 2012 Epistemic complexity and the journeyman-expert transition. *Physical Review Special Topics - Physics Education Research* **8** (1) 010105

[18] Gire E and Manogue C (2012). Making sense of quantum operators, eigenstates and quantum measurements. AIP Conference Proceedings, American Institute of Physics.

[19] Uhden O, et al. 2012 Modelling mathematical reasoning in physics education. *Science & Education* **21** (4) 485-506

[20] Hu D and Rebello N S 2013 Using conceptual blending to describe how students use mathematical integrals in physics. *Physical Review Special Topics - Physics Education Research* **9** (2) 020118

[21] Hu D and Rebello N S 2013 Understanding student use of differentials in physics integration problems. *Physical Review Special Topics-Physics Education Research* **9** 020108

[22] Hu D and Rebello N S 2014 Shifting college students' epistemological framing using hypothetical debate problems. *Physical Review Special Topics - Physics Education Research* **10** (1) 010117

[23] Karam R 2014 Framing the structural role of mathematics in physics lectures: A case study on electromagnetism. *Physical Review Special Topics - Physics Education Research* **10** (1) 010119

[24] Karam R 2015 Introduction of the thematic issue on the interplay of physics and mathematics. *Science & Education* **24** (5) 487-494

[25] Karam R and Krey O 2015 Quod erat demonstrandum: Understanding and explaining equations in physics teacher education. *Science & Education* **24** (5) 661-698

[26] Redish E F and Kuo E 2015 Language of physics, language of math: Disciplinary culture and dynamic epistemology. *Science & Education* **24** 561-590

[27] Tzanakis C (2016). Mathematics & physics: An innermost relationship. Didactical implications for their teaching & learning, Montpellier, France.

[28] Dini V and Hammer D 2017 Case study of a successful learner’s epistemological framings of quantum mechanics. *Physical Review Physics Education Research* **13** (1) 010124

[29] Dreyfus B W, et al. 2017 Mathematical sense-making in quantum mechanics: An initial peek. *Physical Review Physics Education Research* **13** 020141

[30] Odden T O B and Russ R S 2018 Sensemaking epistemic game: A model of student sensemaking processes in introductory physics. *Physical Review Physics Education Research* **14** (2) 020122

[31] Branchetti L, et al. 2019 Interplay between mathematics and physics to catch the nature of a scientific breakthrough: The case of the blackbody. *Physical Review Physics Education Research* **15** (2) 020130

[32] Karam R, et al. (2019). The "math as prerequisite" illusion: Historical considerations and implications for physics teaching. Mathematics in physics education. G. Pospiech, M. Michelini and B.-S. Eylon. Cham, Springer International Publishing**:** 37-52.

[33] Rangkuti M A and Karam R 2022 Conceptual challenges with the graphical representation of the propagation of a pulse in a string. *Physical Review Physics Education Research* **18** (2) 020119

[34] Sirnoorkar A, et al. 2023 Sensemaking and scientific modeling: Intertwined processes analyzed in the context of physics problem solving. *Physical Review Physics Education Research* **19** (1) 010118

[35] Keebaugh C, et al. 2024 Challenges in sensemaking and reasoning in the context of degenerate perturbation theory in quantum mechanics. *Physical Review Physics Education Research* **20** (2) 020139

[36] Reif F 1995 Understanding and teaching important scientific thought processes. *Journal of Science Education and Technology* **4** (4) 261-282

[37] Reif F 1995 Millikan lecture 1994: Understanding and teaching important scientific thought processes. *Am. J. Phys.* **63** (1) 17--32

[38] Reif F 2008 Scientific approaches to science education. *Physics Today* **39** (11) 48--54

[39] Singh C 2002 When physical intuition fails. *Am. J. Phys.* **70** (11) 1103-1109

[40] Leonard W J, et al. 1996 Using qualitative problem-solving strategies to highlight the role of conceptual knowledge in solving problems. *Am. J. Phys.* **64** 1495-1503

[41] Dufresne R, et al. 2005 Knowledge representation and coordination in the transfer process. *Transfer of learning from a modern multidisciplinary perspective* 155--215

[42] Vygotsky L (1978). Mind in society: The development of higher psychological processes, Harvard University Press.

[43] McLeod S 2010 The zone of proximal development and scaffolding. *Recuperado el*

[44] Sweller J 1988 Cognitive load during problem solving: Effects on learning. *Cognitive Science* **12** (2) 257-285

[45] Hutchins E 2014 The cultural ecosystem of human cognition. *Philosophical Psychology* **27** (1) 34-49

[46] Justice P (2019). Helping students learn quantum mechanics using research-validated learning tools. Physics and Astronomy. Pittsburgh, PA 15260, University of Pittsburgh, Ph.D. Dissertation. **Ph.D.**

[47] Justice P, et al. (2022). Impact of mathematical reasoning on students' understanding of quantum optics. Proceedings of the Physics Education Research Conference doi: 10.1119/perc.2022.pr.Justice.

[48] Singh C 2008 Assessing student expertise in introductory physics with isomorphic problems I. Effect of some potential factors on problem solving and transfer. *Physical Review Special Topics - Physics Education Research* **4** (1) 010105

[49] McDermott L C 2001 Oersted medal lecture 2001: "Physics education research—the key to student learning". *Am. J. Phys.* **69** (11) 1127-1137

[50] Kim E and Pak S-J 2002 Students do not overcome conceptual difficulties after solving 1000 traditional problems. *Am. J. Phys.* **70** (7) 759-765

[51] Mazur E (1997). Peer instruction: A user's manual. Upper Saddle River, N.J., Prentice Hall.

[52] Maries A and Singh C 2023 Helping students become proficient problem solvers part i: A brief review. *Education Sciences* **13 (2)** (2) 138

[53] McDermott L C 1991 Millikan lecture 1990: What we teach and what is learned—closing the gap. *Am. J. Phys.* **59** (4) 301-315

[54] Derman A, et al. 2019 Insights into components of prospective science teachers' mental models and their preferred visual representations of atoms. *Education Sciences* **9** (2) 154

[55] Nousiainen M and Koponen I T 2020 Pre-service teachers' declarative knowledge of wave-particle dualism of electrons and photons: Finding lexicons by using network analysis. *Education Sciences* **10** (3) 76

[56] Bitzenbauer P 2021 Quantum physics education research over the last two decades: A bibliometric analysis. *Education Sciences* **11** (11) 699

[57] Chiofalo M L, et al. 2022 Games for teaching/learning quantum mechanics: A pilot study with high-school students. *Education Sciences* **12** (7) 446

[58] Michelini M, et al. 2022 Implementing dirac approach to quantum mechanics in a hungarian secondary school. *Education Sciences* **12** (9) 606

[59] Colletti L 2023 An inclusive approach to teaching quantum mechanics in secondary school. *Education Sciences* **13** (2) 168

[60] Michelini M and Stefanel A (2023). Research studies on learning quantum physics. The International Handbook of Physics Education Research: Learning Physics. M. F. Taşar and P. R. L. Heron. Melville, NY, AIP Publishing LLC Online.

[61] Holincheck N, et al. 2024 Quantum science and technologies in k-12: Supporting teachers to integrate quantum in stem classrooms. *Education Sciences* **14** (3)

[62] Nikolaus P, et al. 2024 Investigating students' conceptual knowledge of quantum physics to improve the teaching and learning process. *Education Sciences* **14** (10) 1113

[63] Doyle L, et al. 2026 Building bridges in quantum information science education: Expert insights to guide framework development for interdisciplinary teaching and evolution of common language. *EPJ Quantum Technology* 13 (2)

[64] Seifollahi F and Singh C 2025 Preparing students for the quantum information revolution: Interdisciplinary teaching, curriculum development, and advising in quantum information science and engineering. *Eur. J. Phys.* **46** (5) 055709

[65] Doyle L, et al. 2026 Do we have a quantum computer? Expert perspectives on current state and future prospects. *Physical Review Physics Education Research* **22** (1) 010101

[66] Singh C 2001 Student understanding of quantum mechanics. *Am. J. Phys.* **69** (8) 885-895

[67] Singh C and Marshman E 2015 Review of student difficulties in upper-level quantum mechanics. *Physical Review Special Topics-Physics Education Research* **11** (2) 020117

[68] Brown B, et al. 2025 Investigating and improving student understanding of time dependence of expectation values in quantum mechanics using an interactive tutorial on larmor precession. *Am. J. Phys.* **93** (1) 52-57

[69] Marshman E and Singh C 2015 Framework for understanding the patterns of student difficulties in quantum mechanics. *Physical Review Special Topics-Physics Education Research* **11** (2) 020119

[70] Galvez E J, et al. 2005 Interference with correlated photons: Five quantum mechanics experiments for undergraduates. *Am. J. Phys.* **73** 127

[71] Beck M (2012). Quantum mechanics: Theory and experiment, Oxford University Press.

[72] Orszag M (2016). Quantum optics: Including noise reduction, trapped ions, quantum trajectories, and decoherence, Springer.

[73] Scholz R, et al. 2018 Undergraduate quantum optics: Experimental steps to quantum physics. *Eur. J. Phys.* **39** (5) 055301

[74] Galvez E J (2019). Quantum optics laboratories for teaching quantum physics. Education and Training in Optics and Photonics, Optica Publishing Group.

[75] Bitzenbauer P and Meyn J-P 2020 A new teaching concept on quantum physics in secondary schools. *Physics Education* **55** (5) 055031

[76] Blais A, et al. 2020 Quantum information processing and quantum optics with circuit quantum electrodynamics. *Nature Physics* **16** (3) 247-256

[77] Fox M F J, et al. 2020 Preparing for the quantum revolution: What is the role of higher education? *Physical Review Physics Education Research* **16** (2) 020131

[78] Bitzenbauer P 2021 Practitioners' views on new teaching material for introducing quantum optics in secondary schools. *Physics Education* **56** (5) 055008

[79] Bitzenbauer P 2021 Development of a test instrument to investigate secondary school students' declarative knowledge of quantum optics. *European Journal of Science and Mathematics Education* **9** (3) 57-79

[80] Singh C, et al. (2021). Preparing students to be leaders of the quantum information revolution. Physics Today DOI: 10.1063/PT.6.5.20210927a https://physicstoday.scitation.org/do/10.1063/PT.6.5.20210927a/full/

[81] Asfaw A, et al. 2022 Building a quantum engineering undergraduate program. *IEEE Transactions on Education* **65** (2) 220-242

[82] Bitzenbauer P, et al. 2022 Assessing engineering students' conceptual understanding of introductory quantum optics. *Physics* **4** (4) 1180-1201

[83] Posner M T, et al. 2022 Special section guest editorial: Education and training in quantum sciences and technologies. *Optical Engineering* **61** (8) 081801-081801

[84] Seskir Z C, et al. 2022 Quantum games and interactive tools for quantum technologies outreach and education. *Optical Engineering* **61** (8) 081809-081809

[85] Walsh J A, et al. 2022 Piloting a full-year, optics-based high school course on quantum computing. *Physics Education* **57** (2) 025010

[86] Borish V and Lewandowski H J 2023 Seeing quantum effects in experiments. *Physical Review Physics Education Research* **19** (2) 020144

[87] Galvez E J (2023). A curriculum of table-top quantum optics experiments to teach quantum physics. Journal of Physics: Conference Series, IOP Publishing.

[88] Hasanovic M (2023). Quantum education: How to teach a subject that nobody fully understands. Education and Training in Optics and Photonics, Optica Publishing Group.

[89] Nita L, et al. 2023 The challenge and opportunities of quantum literacy for future education and transdisciplinary problem-solving. *Research in Science & Technological Education* **41** (2) 564-580

[90] Silberman D M (2023). Teaching quantum to high school students. Education and Training in Optics and Photonics, Optica Publishing Group.

[91] Ubben M and Bitzenbauer P 2023 Exploring the relationship between students' conceptual understanding and model thinking in quantum optics. *Frontiers in Quantum Science and Technology* **2** 1207619

[92] Hennig F, et al. 2024 A new teaching-learning sequence to promote secondary school students' learning of quantum physics using dirac notation. *Physics Education* **59** (4) 045007

[93] Hennig F, et al. 2024 Mathematical sense making of quantum phenomena using dirac notation: Its effect on secondary school students' functional thinking about photons. *EPJ Quantum Technology* **11** (1) 61

[94] Tóth K, et al. 2024 Exploring the effect of a phenomenological teaching-learning sequence on lower secondary school students' views of light polarisation. *Physics Education* **59** (3) 035009

[95] Marshman E and Singh C 2017 Investigating and improving student understanding of quantum mechanics in the context of single photon interference. *Physical Review Special Topics-Physics Education Research* **13** (1) 010117

[96] Marshman E and Singh C 2016 Interactive tutorial to improve student understanding of single photon experiments involving a mach–zehnder interferometer. *Eur. J. Phys.* **37** (2) 024001

[97] Marshman E and Singh C (2022). QuILTs: Validated teaching–learning sequences for helping students learn quantum mechanics. Physics teacher education: What matters? J. Borg Marks, P. Galea, S. Gatt and D. Sands. Cham, Springer International Publishing**:** 15-35 https://doi.org/10.1007/978-3-031-06193-6_2 .

[98] Singh C, et al. (2023). Instructional strategies that foster effective problem-solving. The International Handbook of Physics Education Research: Learning Physics, Melville, New York, AIP Publishing LLC https://aip.scitation.org/doi/pdf/10.1063/9780735425477_017

[99] Henderson C and Beichner R 2009 Publishing per articles in ajp and prst-per. *Am. J. Phys.* **77** (7) 581-582

[100] Schwartz D L, et al. 2005 Efficiency and innovation in transfer. *Transfer of learning from a modern multidisciplinary perspective* **3** (1) 1-51

[101] Rebello N S, et al. (2017). Transfer of learning in problem solving in the context of mathematics and physics. Learning to solve complex scientific problems. D. H. Jonassen, Routledge**:** 223-246.

[102] Maries A, et al. 2020 Can students apply the concept of "which-path" information learned in the context of mach–zehnder interferometer to the double-slit experiment? *Am. J. Phys.* **88** (7) 542-550

[103] Physport. "Physport." from https://www.physport.org/.

[104] Hake R R 1998 Interactive-engagement versus traditional methods: A six-thousand-student survey of mechanics test data for introductory physics courses. *Am. J. Phys.* **66** (1) 64-74

[105] Cohen J (2013). Statistical power analysis for the behavioral sciences, Routledge.